\documentclass[amssymb,prl,epsfig,superscriptaddress,preprint,authortitle ]{revtex4-2}
\usepackage[utf8]{inputenc}
\usepackage{color}
\usepackage[normalem]{ulem}
 
\usepackage{amssymb}

\usepackage{graphicx}
\usepackage{epsfig}
\usepackage{amssymb}
\usepackage[pdftex, bookmarks,colorlinks=true,urlcolor=blue]{hyperref}

\begin{document}
\title{Tracking defects of Electronic Crystals by Coherent X-ray Diffraction}

\author{D. Le Bolloc'h\footnote{Corresponding author: david.le-bolloch@universite-paris-saclay.fr}}
\affiliation{Université Paris-Saclay, CNRS, Laboratoire de Physique des Solides, 91405, Orsay, France}

\author{E. Bellec}
\affiliation{ESRF - The European Synchrotron, 71 Avenue des Martyrs, F-38000 Grenoble, France}

\author{N. Kirova\footnote{Corresponding author: natacha.kirova@universite-paris-saclay.fr}}
\affiliation{Université Paris-Saclay, CNRS, Laboratoire de Physique des Solides, 91405, Orsay, France}

\author{V.L.R. Jacques}
\affiliation{Université Paris-Saclay, CNRS, Laboratoire de Physique des Solides, 91405, Orsay, France}

\begin{abstract}

In this article, we review different studies based on advanced x-ray diffraction techniques - especially coherent x-ray diffraction - that allowed us to reveal the behaviour of such symmetry-breaking systems as Charge Density Wave (CDW) and Spin density Wave (SDW), through their local phase. After a brief introduction on the added value of using coherent x-rays, we show how the method can be applied to CDW and SDW systems, in both static and dynamical regimes. The approach allowed us to probe the particular sliding state of  CDWs systems by observing them through their phase fluctuations, to which coherent x-rays are particularly sensitive. Several compounds stabilizing a CDW phase able to slide are presented, each with a different but clearly pronounced signature of the sliding state. Two main features emerge from this series of experiments which have been little treated until now, the influence of CDW pinning by the sample surfaces and the propagation of periodic phase defects such as charge solitons across the entire sample. Phase models describing the spatial and temporal properties of sliding CDWs are presented in the last part of this review.

\end{abstract}

\maketitle
\section{Introduction}
As early as in the 1970s, several authors raised the importance of the phase in CDW systems \cite{LEE1974703,Fukuyama1978}. Indeed, a Charge Density Wave (CDW) is described by a periodic modulation of charges $\rho(\vec r)=A\cos [2k_{F}\vec r+ \phi(\vec r)]$, where A is the amplitude and $\phi$ is the phase which denotes the position of the CDW relative to the atomic host lattice. As a matter of facts, external perturbations generally mainly affect the CDW phase. For instance, when submitting the system to an electric current, the threshold field above which the CDW {\it depins} from the atomic lattice and slides, leading to an additional current, is directly linked to local CDW phase variations, either through defects in the bulk \cite{Fukuyama1978}, conversion processes at the electrodes \cite{Ong:85om} or pinning at the surface \cite{PhysRevLett.70.845,Rideau_2002}. Although CDW deformation and phase shifts have been theoretically studied for a long time \cite{Feinberg}, the precise observation of the phase deformation has been missing until the advent of advanced x-ray diffraction techniques. 

From a general point of view, the  observation of all types of defects in condensed matter has always been challenging. Electron diffraction methods on thin samples or surface techniques such as STM are very efficient to observe crystal dislocations at the atomic scale. On the other hand, bulk experiments such as neutron or x-ray diffraction are also sensitive to defects but result from spatial averages which provide a global view of disorder at the macroscopic scale.
Observing  localized CDW phase shifts is a much more difficult task. In that case indeed, the phase shift does not concern the host lattice itself, but the periodic atomic displacement associated to the CDW, small in amplitude and which overlaps to the host atomic lattice. To some extent, this type of defect could be called  a second order phase shift. The purpose of this review is to show how coherent x-ray diffraction can provide access to such peculiar phase singularities. 

Experiments using coherent x-ray beams have been being developed continuously since the 90's thanks to the improvement of synchrotron sources \cite{sutton1991}. Third generation sources are indeed able to deliver much brighter beams and smaller source sizes, allowing to take advantage of the coherence properties of the beam.

After presenting the methodological aspects from model examples, we will show how this technique has proved efficient to probe CDW phase shifts and their dynamics. 
    
    \section{Coherent X-ray diffraction to track phase shifts in condensed matter}
    
Speaking of coherent diffraction is actually a pleonasm. Indeed, the diffraction process originates from contructive interferences and is therefore a coherent phenomenon in essence. However, this expression is justified by the orders of magnitude involved. Indeed, the beam is defined by two characteristic lengths: the first one is related to the wavelength, and the second one to the relative angle of propagation. We thus define two coherence lengths, the longitudinal coherence length $\xi_l = \frac{\lambda^2}{2\Delta\lambda}$ and the transverse one $\xi_t = \frac{\lambda R}{2a}$, where $\lambda$ is the beam wavelength, $\Delta\lambda$ the spectral width of the source, $a$ is the numerical aperture of the source and R is the distance from the source. These two quantities have to be large enough to see interferences. But how large? It actually depends on the typical size of the object under consideration. For instance, in the case of classical x-ray diffraction on crystals,  $\xi_l$ and $\xi_t$ have to be larger than the lattice parameters of the chosen crystal, which is always the case, even for laboratory x-ray sources. However, to obtain interference from larger objects, both coherence lengths must be scaled to the dimensions of the object. This is standard to get interferences from micron-size objects with visible light, using lasers typically, but harder to get with x-rays as $\xi_l$ and $\xi_t$ scale with $\lambda$. However, since the emergence of third-generation synchrotrons, micron-size values for $\xi_t$ can be obtained thanks to micrometer source sizes and large distances between source and sample, while $\xi_l$ in the micron range is achieved thanks to low bandwidth monochromators. 
    Hence, we generally speak of coherent diffraction, when the coherence lengths of the x-ray beam are close to the size of the diffracted entities.
    
\begin{figure}[!ht]
    \centering
    \includegraphics[width=5in]{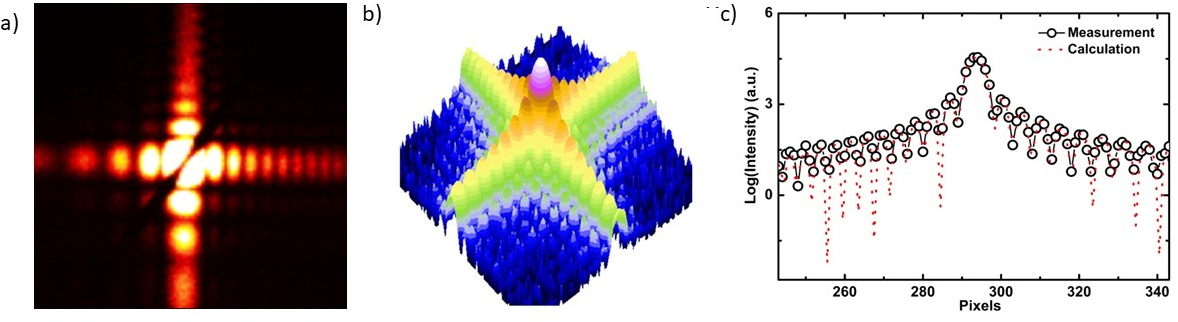}
    \caption{a) and b) Coherent diffraction pattern of a 5$\mu$m square slit using a 7keV x-ray beam. c) Profile of the diffraction pattern showing high-visibility fringes.}
    \label{fig:fig1}
    \end{figure}

    The diffraction pattern of a rectangular slit opened at few micrometers and leading to the expected cardinal sinus squared diffraction pattern (see Fig.~\ref{fig:fig1}) is an illustration the phenomenon. The very good contrast of the interference fringes reveals the high degree of coherence obtained in the hard x-ray regime~\cite{LeBolloch2002,Jacques2012a,lebolloch2011}. 
    
    {}
    
     Many techniques emerged to take advantage of the coherent properties of beams produced by large-scale instruments, all based on the analysis of the interference patterns obtained by objects introducing a phase shift in the beam propagation. In condensed matter physics, any system that deviates from perfect crystallinity and/or is smaller than the beam size introduces such phase shifts and leads to interference patterns. Coherent x-ray diffraction (CXRD) thus opened the way to new opportunities, as the possibility to follow the fluctuation dynamics in condensed matter (called x-ray photon correlation spectroscopy) or to obtain the real-space image of the diffracted object using phase-retrieval methods (referred to as coherent diffraction imaging methods \cite{Robinson2009,Chapman2010}).
     
     There is however another possible application of the use of a coherent beam. Among all the possible defects encountered in condensed matter, some of them introduce phase shifts, as dislocations, for which coherent diffraction is particularly sensitive.  Before discussing the case of dislocations in electronic crystals, let us illustrate the phenomenon with the textbook case of an isolated dislocation in a perfect crystal.
     
\begin{figure}[!ht]
    \centering
    \includegraphics[width=3.37in]{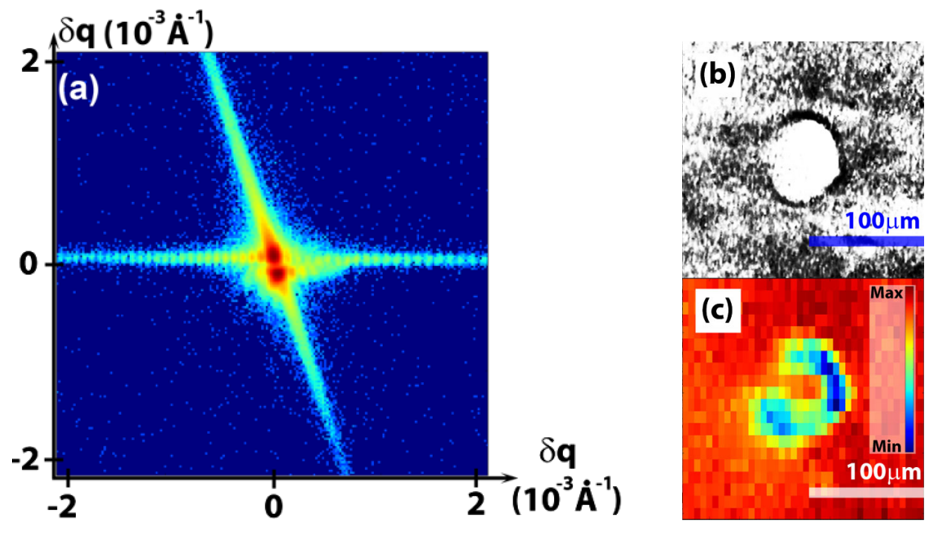}
    \caption{Coherent diffraction of a Silicon crystal displaying a dislocation loop a) CXRD pattern obtained on the 220 reflection when the beam probes the dislocation line. (b) Topography image of a typical dislocation loop in this sample. (c) Image obtained in the same area as (b) by scanning the area with the 5$\mu$m coherent x-ray beam and plotting the intensity at the expected Bragg peak position. This experiment has been performed at the CRISTAL beamline of the SOLEIL synchrotron at E=7keV.}
    \label{fig:fig2}
    \end{figure}

     For example, an isolated dislocation loop can be stabilized in a silicon crystal after specific thermal treatments. Such a defect introduces phase-shifted domains on each side of the dislocation line. When a coherent x-ray beam probes regions containing such dislocation lines, interferences are observed (see Fig.~\ref{fig:fig2}). 
     In perfects regions of the crystal, a well defined Bragg peak is observed. In contrast, when the beam probes the dislocation line, a destructive interference is observed, and the Bragg peak displays two side maxima (Fig.~\ref{fig:fig2}.a). The local minimum in between the two maxima can be tracked as a function of beam position on the sample to retrieve the full dislocation loop (Fig.~\ref{fig:fig2}.c). The resulting image is in agreement with images obtained by x-ray topography  (Fig.~\ref{fig:fig2}.b). In addition, more details can be extracted of the CXRD pattern, especially from the oblique scattering line that reveals that these dislocation loops are dissociated into two partials (see~\cite{Jacques2011} for more details).
    
    \subsection{Phase shifts of electronic crystals studied by coherent X-ray diffraction}
    The same methodology can be applied to electronic crystals (CDW and SDW) focusing the measurement on the satellite reflection associated to the periodic lattice distortion instead of the diffraction peaks associated to the host atomic lattice itself.
    
    \subsection{Isolated CDW and SDW dislocations}
\begin{figure}[!ht]
    \centering
    \includegraphics[width=3.5in]{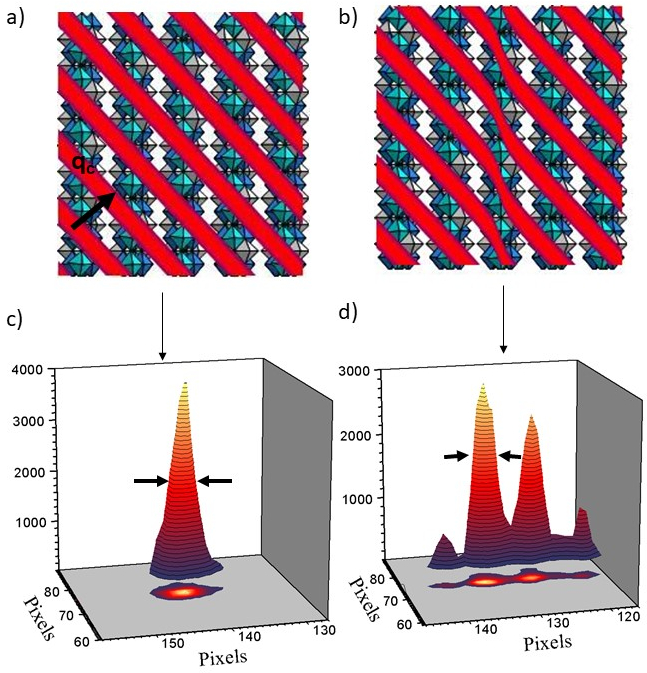}
    \caption{Schematic representation of the CDW
wavefronts in the blue bronze (red lines) in the quasi-1D crystal structure (only MoO$_6$ octahedra are represented): a) without a CDW dislocation and b) with a CDW mixed-dislocation. Coherent diffraction pattern measured at two different sample positions at T = 75K. The single peak measured in c) corresponds to a perfect CDW as represented in a) whereas interference fringes in d) are consistent with a CDW displaying a mixed dislocation, between an edge and a screw dislocation, as displayed in b). Note that the widths of the two fringes are identical and equal to the width of the reflection associated to the perfect CDW.}
    \label{fig:fig3}
    \end{figure}

    First, electronic crystals can display their own phase defects such as dislocations that can be probed with a coherent beam, as described in the previous section. A first demonstration of a CDW dislocation was obtained in the blue bronze K$_{0.3}$MoO$_3$ \cite{Bolloch2005}. This compound is a quasi-1D system made of chains of MoO$_6$ clusters along which an incommensurate CDW develops below 183K. The CDW being out-of-phase for adjacent planes, the CDW wavefronts are inclined with respect to the chain direction, as illustrated in Fig.~\ref{fig:fig3}.a. 
    
    In most regions of the sample, the satellite reflection associated to the CDW modulation displays a single peak (Fig.~\ref{fig:fig3}.c), indicating a long-range order greater than the beam size without CDW phase defect in the micron-sized probed volume. In other regions however, the CXRD pattern is split into two subpeaks with the same widths (Fig.~\ref{fig:fig3}.d). Similarly to the diffraction of a slit where all fringes have the same width (at half maximum intensity), all fringes display here the same widths. This is the typical signature of interference effects between two domains out of phase. This diffraction profile can be well reproduced by considering a mixed-dislocation of the CDW, between an edge and a screw dislocation,  as schematically shown in Fig.~\ref{fig:fig3}.b).
    
\begin{figure}[!ht]
    \centering
    \includegraphics[width=6in]{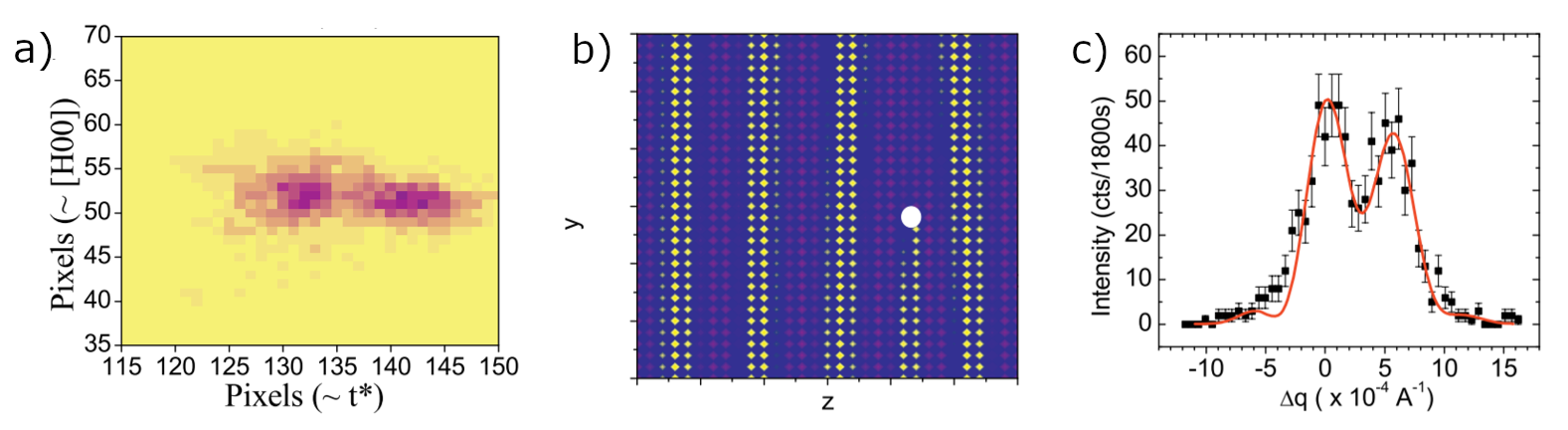}
    \caption{a) CXRD pattern obtained on the SDW peak of chromium at a position where the peak is split, as seen on the CCD detector; b) simulation of an edge dislocation line of the SDW. The white dot is the dislocation line position, and the yellow/red dots are a representation of the magnetic moments at each atomic position in the presence of a SDW dislocation; c) red solid line: simulated CXRD pattern corresponding to the spatial arrangement shown in b) represented on top of the experimental data (black squares).}
    \label{fig:sdw_dislo}
    \end{figure}

    Similar isolated phase defects can be detected in magnetic modulations, like SDW. This was observed in chromium, that exhibits both SDW and CDW modulations below T$_N$ = 311K with associated satellites reflections at wave vectors q$_s$=2k$_F$ and q$_c$=4k$_F$ respectively. Using CXRD in a non-resonant magnetic mode, a characteristic splitting of the 2k$_F$ satellite reflection associated to the SDW is observed at certain positions of the sample, while a single peak is visible at most other positions, and turned out to be in agreement with an edge-dislocation on the magnetic modulation (see Fig.~\ref{fig:sdw_dislo}) \cite{Jacques2009b}. 
    
     \subsection{Coexisting SDW and CDW: two modulations with highly different correlation lengths}
      Coherent diffraction experiments also allow  to compare the state of disorder through speckles. Indeed, the number of speckles observed is, in first approximation, related to the number of defects. This property is particularly interesting in the case of chromium case that hosts two coexisting phases whose type of interaction has been much discussed.
       \begin{figure}[!ht]
    \centering
    \includegraphics[width=5in]{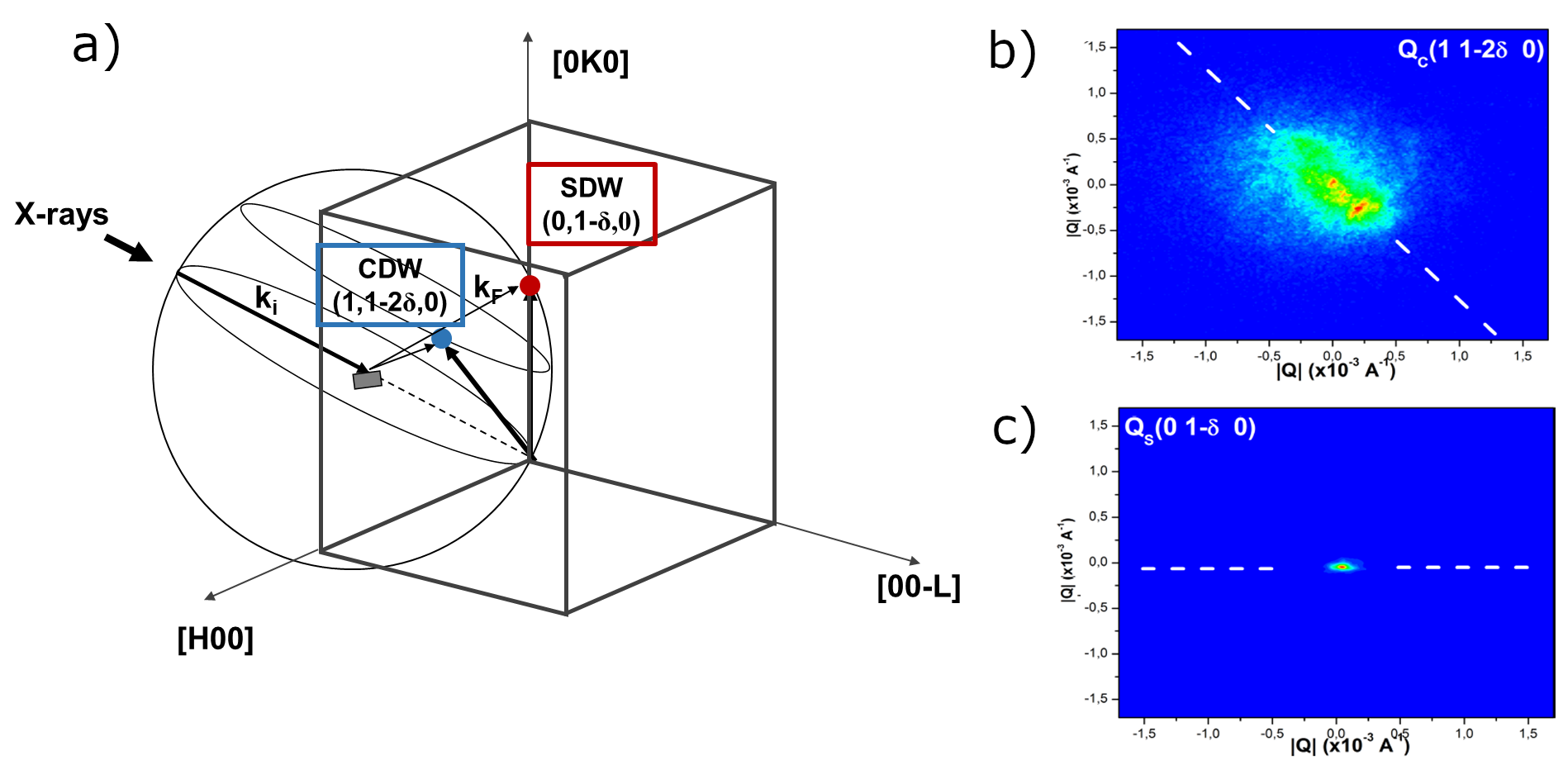}
    \caption{a) Schematic drawing of the simultaneous diffraction experiment. Given an incident wave vector $k_i$, the sample is oriented so that both $q_S$ and $q_C$ satellite reflections are simultaneously located on the Ewald sphere, and therefore both fulfill the diffraction condition. 
Comparison of b) CDW and c) SDW satellite reflections in chromium by using
coherent and simultaneous diffraction through the maximum intensity of the CDW $(q_{C}{=}(1{,}1{-}2\delta,0))$ and SDW satellites $(q_{S}{=}(0{,}1{-}\delta,0))$ (beam size=10$\mu m$$\times$10$\mu m$). For most of the regions probed, the $q_C$ satellite displays speckles, while no speckle is observed at $q_{S}$.}
    \label{fig:fig4}
    \end{figure}

    If the SDW represents the main harmonics of the modulation with wavevector $\vec{Q}$, a CDW is concommitantly stabilized as its second harmonics at $\vec{2Q}$. However, while the magnetic instability is clearly due to the nesting of electron and hole pockets at the Fermi level with wave vector $\vec{Q}$, the origin of the CDW is far less understood. Several scenarii may account for its appearance. The first one relies on magnetostrictive coupling : the interaction between the SDW and the atomic lattice induces a CDW at $\vec{2Q}$~\cite{Cowan1978}. Another hypothesis involves a second nesting of unnested hole pockets following the SDW formation~\cite{Young1974}. How to distinguish between this purely electronic or magneto-elastic scenarii? As coherent diffraction is very sensitive to local defects, similar profiles on the two reflections are expected in the case of the magnetostrictive origin of the CDW. However, comparing the singularities of the two phases is not an easy task in diffraction because the probed volumes are in general not equivalent. The way to avoid this is to use simultaneous diffraction geometry for the SDW and CDW reflections by placing simultaneously the $q_S=(0,1-\delta,0)$ SDW reflection and the $q_C=(1,1-2\delta,0)$ CDW reflection on the Ewald sphere (see Fig.~\ref{fig:fig4}.a). The images recorded at the maximum of the rocking curve for the CDW and SDW reflections using a 2D detector are displayed in Fig.~\ref{fig:fig4}.b) and c) respectively, with the same scale in reciprocal space. 
   
The difference between the two diffraction patterns is striking: while the CDW reflection is broad and contains many speckles, the SDW reflection is as narrow as the direct beam. This reveals a high number of CDW defects while the SDW correlation length remains larger than the 10$\mu m$*10$\mu m$ probed volume. We can directly infer from these measurements that the the origin of the CDW is not directly linked to that of the SDW. The scenario relying on a purely magnetostrictive origin of the CDW does not hold while the results could be compatible with a band model based on a second order nesting~\cite{Jacques2014}.

    \section{Dynamics of CDW sliding based on phase shifts motion}

The most spectacular property of incommensurate CDW systems is their ability to carry a collective current when submitting the sample to an external electric field (for a review, see \cite{Gork'ov1989,Monceau2012}).  Above a threshold bias current, an oscillating current is detected with a fundamental frequency as well as several harmonics \cite{Fleming:1979cu}. Up to 23 harmonics have been observed in NbSe$_3$ \cite{thorne:1987bt}. This collective transport of charges through macroscopic sample has received a considerable interest for more than 35 years~\cite{gruner2018density,Gork'ov1989}. However, the understanding of the type of charge carriers involved in the phenomenon and their propagation mode still remains incomplete.

The first proposed scenario was based on the translational invariance of the incommensurate modulation allowing the whole CDW to slide over the atomic lattice without dispersion \cite{Frohlich:1954fz}. In fact, as we will see in the following, the sliding state is characterized by a strong distortion of the CDW. In addition, the CDW is an almost sinusoidal modulation as shown by diffraction experiments (the second harmonic is usually very weak in intensity) while transport measurements reveal a strong anharmonic signal \cite{thorne:1987bt} which suggests that the transport of charges in CDW systems is far more complex than a simple CDW translation. 
Another description considering the influence of defects, assumes a slowly varying phase $\phi(x)$ of the CDW interacting weakly with impurities \cite{Fukuyama1978}. The existence of the threshold field is thus well explained by considering an empirical bulk pinning potential, either strong or weak, depending on the type of defects and their concentration \cite{gruner2018density}. On the other hands, a strong electron-phonon coupling has been considered where the lattice itself plays the role of CDW pinning leading to discontinuities, in the phonon spectrum and atomic modulation \cite{aubry1989}. A pure quantum tunneling through the sample was also proposed \cite{Bardeen:1979hv}. However, 
the most accepted theory, developed by Ong and Maki \cite{Ong:85om} and Gor'kov \cite{Gorkov:1983tb, Gorkov:1984vj, Batistic:1984by}, deals with the CDW-metal junction at  electrical contacts. The conversion of normal electrons from the metallic electrode into  condensed charges in the CDW is made possible by climbing CDW dislocations at the interface.
This so-called phase slippage and current conversion phenomena are in agreement with local resistivity measurements  close to contacts \cite{maher1995}. In the phase slippage theories \cite{Ong:85om}, impurities play a minor role, hidden in the tunneling coefficient. Note also that phase slippage mixed with quantum tunneling have been considered at low temperature \cite{MAKI1995}. 

The validation of either of these theories suffers, however, from the fact that it is very difficult to observe this phenomenon at the atomic scale. The aim of this review is to show how the use of CXRD to observe CDW phase defects brings new insight on charge transport in CDW materials.

  \subsection{Dynamics of sliding CDW revealed by CXRD}
  
The sliding state was historically observed by macroscopic resistivity measurements, but its signature in diffraction is also clear.
  Although each CDW system displays its own behavior, the sliding state is characterized  in all cases by an increase of disorder below the threshold. Depending on the system under study, the type of disorder may take the form of creep, compression, expansion, rotation, or shear of the CDW wavefronts, with ordering processes by motion for larger currents or the appearance of an additional modulation appearing on top of the CDW. 

\begin{figure}[!ht]
    \centering
    \includegraphics[width=6in]{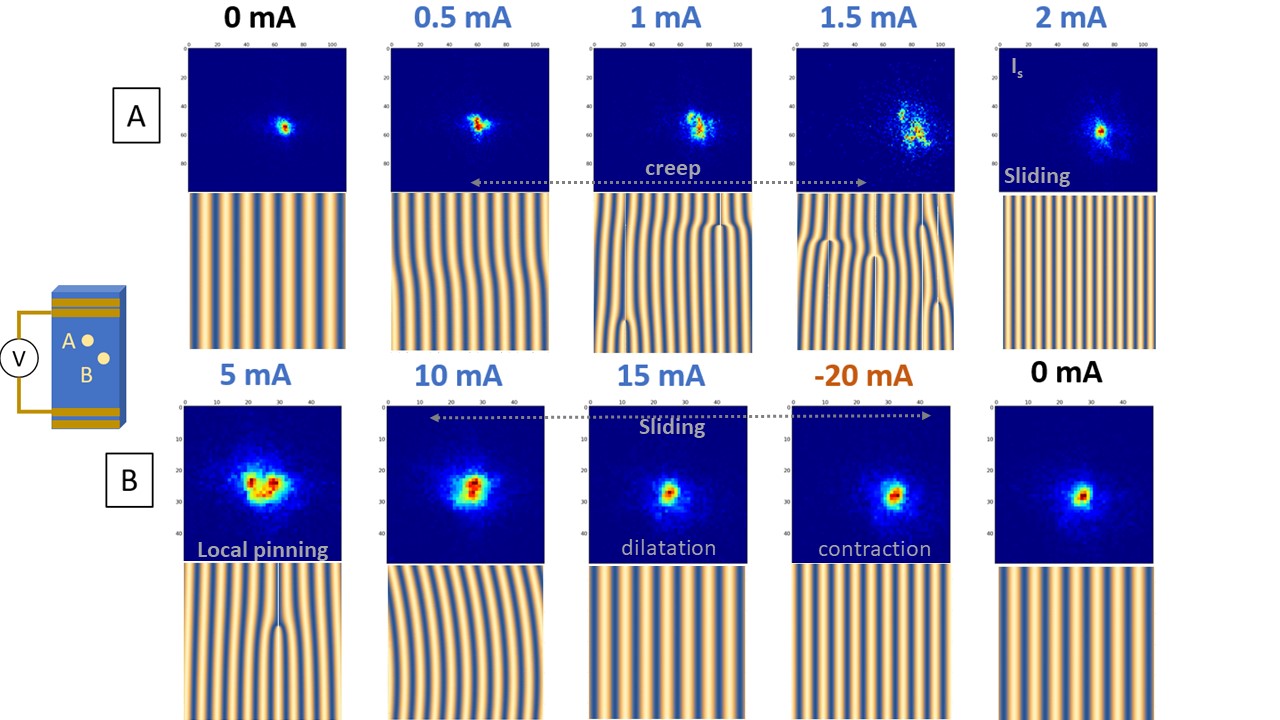}
    \caption{Coherent diffraction patterns of the 2k$_F$ satellite reflection associated to the CDW of the blue bronze, at two positions A and B, while applying external dc currents. Each diffraction pattern corresponds to a sum of the full rocking curve and has been obtained for currents between 0mA and 2mA at position A and from 5mA to 15mA, then switched to -20mA and back to 0mA at position B. A sketch of the corresponding wavefront configuration in real space is displayed below each image, illustrating the creep regime below the threshold, the narrowing effect above the threshold due to the sliding motion, as well as dilatation and contraction of the CDW wavelength in the sliding state depending on the current direction. This experiment was performed at the CRISTAL beamline of SOLEIL synchrotron.}
    \label{fig:fig5}
    \end{figure}

As an illustration of the diversity of the phenomenon, let us first describe the behavior of the blue bronze K$_{0.3}$MoO$_3$ system under current (see Fig.~\ref{fig:fig5}).
    {} 
In this experiment, an external current was applied to the sample in a 4-probe configuration at 70K, below and above the threshold current $I_S=2$mA. In most regions of the sample, the behaviour is similar to the one measured at position A in Fig.~\ref{fig:fig5}: in the virgin state (I=0mA), the CDW reflection is made of a single peak, which accounts for a long-range order of the CDW. When current is applied, the CDW reflection broadens and displays speckles, which shows that the CDW loses its coherence even at very low currents, far below the threshold $I_S$, which is characteristic of the creep regime. Above $I_S$, when the macroscopic excess of current is observed by transport measurements, the CDW reflection gets narrower and accounts for a recovery of long-range order. However, this observation remains very local and is not homogeneous from one place to another. At another beam position (the beam size is few microns large) some intrinsic defects locally pin the CDW even above $I_S$, which gives rise to speckles (at 5mA here, see Fig.~\ref{fig:fig5}b). If current is further increased, the CDW can overcome the pinning center and recover its long-range order, here at I=15mA. By reversing the current, at I=-20mA, this long-range order is maintained, and still apparent when the current is switched off~\cite{PhysRevLett.100.096403,PhysRevB.85.035113}.
 
\begin{figure}[!ht]
    \centering
    \includegraphics[width=4.5in]{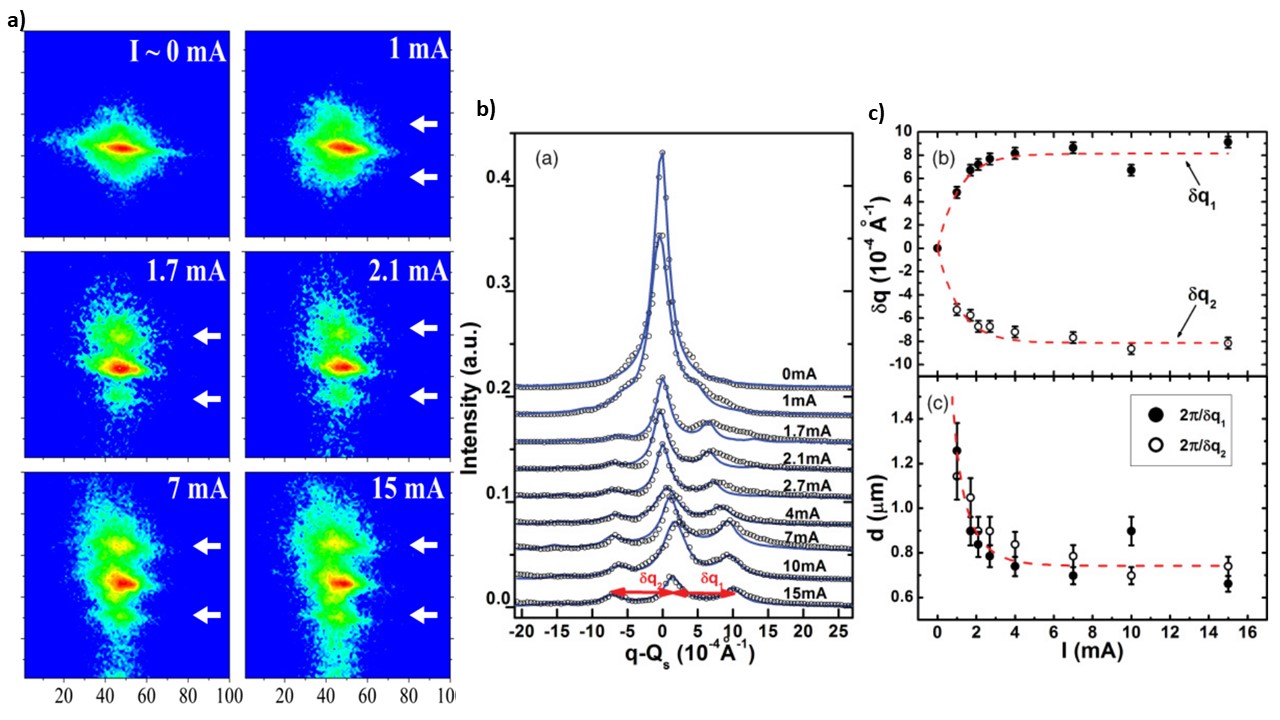}
    \caption{(a) 2D maps of the CDW satellite summed over the full rocking curve versus applied currents and (b) corresponding CDW profile along $\vec{b^*}$ (sum over the vertical CCD axis). c) Evolution of the two secondary satellites position with respect to the 2k$_F$ position (top panel) and the corresponding period in real space reaching more than one micrometer (lower panel).}
    \label{fig:secondary}
    \end{figure}

Finer details revealing the effect of sliding can be detected. Probing the 2k$_F$ CDW satellite with respect to external $dc$ currents in the blue
bronze reveals the existence of an extra modulation (see Fig.~\ref{fig:secondary}). 

  {}

In the sliding regime, the 2k$_F$ satellite reflection displays secondary satellites along the chain axis which corresponds to the appearance of a new periodicity in the system, with periods in the micrometer scale \textit{i.e.} 1500 times larger than the CDW wavelength that decrease with increasing currents (see Fig.~\ref{fig:secondary}). We will come back to this experiment in the next chapter.

\begin{figure}[!ht]
    \centering
    \includegraphics[width=2.8in]{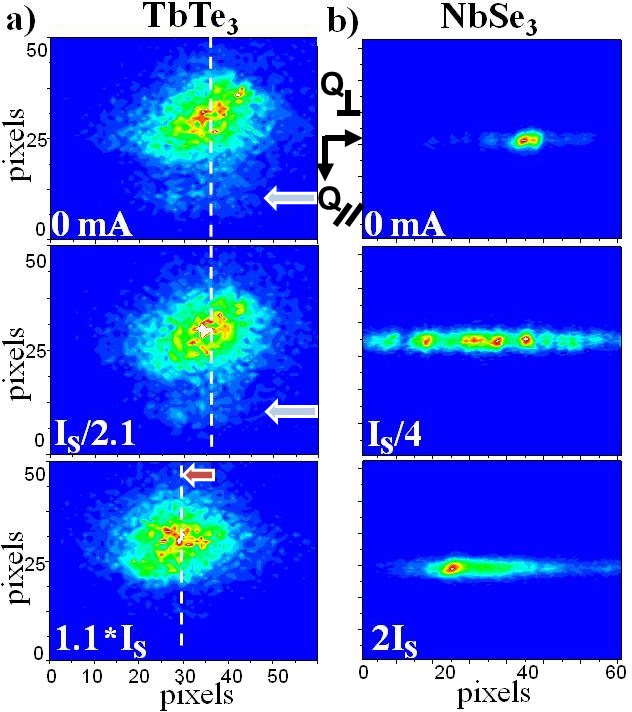}
    \caption{Coherent diffraction patterns of the $2k_F$ satellite reflection associated to the CDW versus external current, below and above the threshold current I$_S$, in a) the quasi-two dimensional TbTe$_3$ system (for I=0mA, I=I$_S$/2.1=5mA and I=1.1$\times$ I$_S$=12mA)  where the red arrow indicates the shift of the 2k$_F$ reflection at I$_S$. b) in the quasi-one dimensional NbSe$_3$ system displaying an elongated shape made of speckles below the threshold before refining above. The 2D images are a sum over several incidence angles through the maximum of intensity.  Although the cutting plane is different in the two cases, the vertical  direction of the camera  is close to the 2k$_F$ wave vector  (Q$_\parallel$) and the horizontal one is transverse to  2k$_F$ (Q$_\perp$) in both cases \cite{PhysRevB.93.165124}.}
    \label{fig:tbte3}
    \end{figure}

The sliding state is characterized by different features depending on the system under consideration. Fig.~\ref{fig:tbte3} shows a comparison between two other sliding systems: the quasi two dimensional system TbTe$_3$ and the quasi one dimensional one NbSe$_3$, both probed by CXRD. Although the two systems do not display the same behavior, the diffraction patterns are very sensitive to the threshold current in both cases.  

TbTe$_3$ samples are intrinsically much less ordered than the blue bronze or NbSe$_3$, leading to broad diffraction peaks and speckles.
However, this disorder does not prevent the system from sliding \cite{PhysRevB.85.241104} and the non-Ohmic conductivity is intimately linked to a strong distortion of the CDW.  

The satellite reflection associated to the CDW shown in Fig. 8.a) displays speckles even without current. Below the threshold current I$_s$, the peak remains unchanged,
but a visible shift in position is observed above I$_s$. This global shift corresponds to a 
rotation of the CDW wave vector in the sliding state (Fig.~ \ref{fig:tbte3}.a) \cite{PhysRevB.93.165124}. Despite this reordering, one can still observe speckles surrounding the peak proving that the CDW remains in a disordered state above I$_s$.

{}

The case of NbSe$_3$ is quite different. The diffraction pattern without current displays an almost single peak corresponding to CDW correlation lengths larger than the beam size, \textit{i.e.} more than several micrometers in all directions. For small currents, far below the threshold, the satellite reflection displays an elongated shape along the transverse direction and is made of speckles (see Fig.~\ref{fig:tbte3}.b).
   \begin{figure}[!ht]
    \centering
    \includegraphics[width=4.37in]{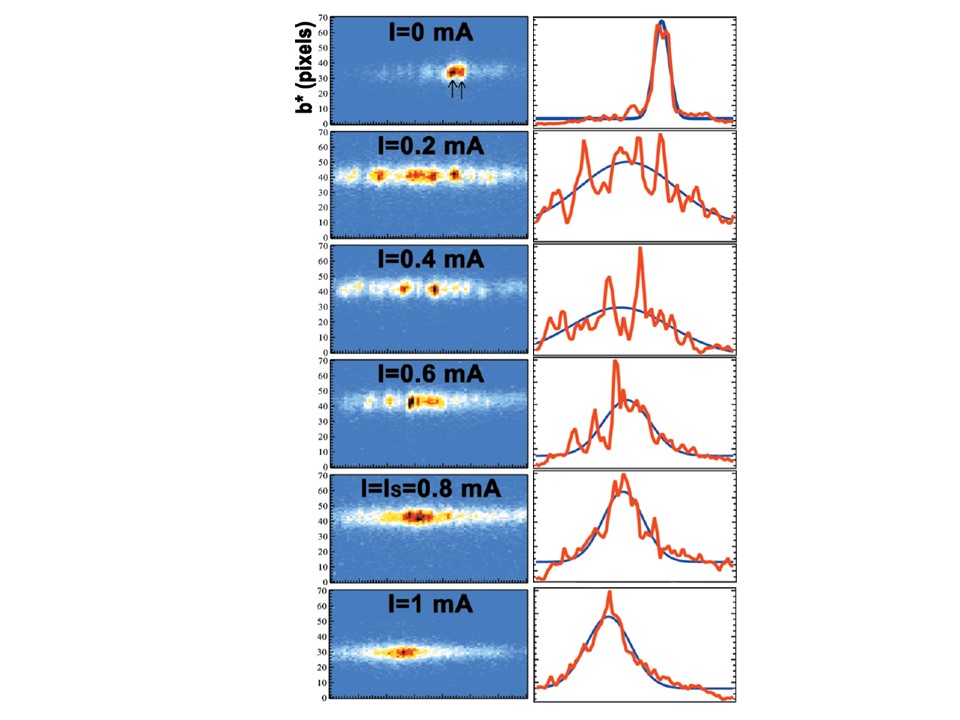}
    \caption{2D diffraction patterns of the
(0,1.241,0) satellite reflection under external current from
0.2 to 1.8 mA (left column), as well as the corresponding
transverse profiles (right column) obtained after integration over the longitudinal direction. Speckles are observed even for weak currents and disappear above the threshold at I$_s$=0.8mA due to time average~\cite{PhysRevLett.109.256402}.}
    \label{fig:nbse3}
    \end{figure}{}
  
Once the applied current exceeds the threshold current for sliding, speckles disappear in NbSe$_3$ leading to smooth diffraction profiles (see Fig.~ \ref{fig:tbte3}b). This effect is more visible in Fig.~\ref{fig:nbse3}. The disappearance of speckles does not correspond to a decreasing number of CDW phase shifts but to time average, the counting time to get the image being longer than the characteristic time of CDW sliding. Indeed, the 10s acquisition time necessary to obtain
one diffraction pattern is long compared to the phase shifts motion~\cite{PhysRevLett.109.256402}.

  \subsection{Microscale shear deformation of a CDW induced by surface pinning} 
    
    The limitation of the previous experiment is that the probe remains static and large, averaging the CDW over the whole illuminated region of the sample, and thus excluding the possibility to observe local variations of the CDW.
The NbSe$_3$ system, however, is known to develop a continuous CDW deformation under current. Indeed,
    current-induced CDW deformations have been measured mainly close to the two electrodes by using a 50$\mu m\times$50$\mu m$ x-ray beam along the entire sample length. The CDW appears to be compressed on one edge close to the electrical contact and expanded on the other, leading to a clear phase asymmetry in this direction  \cite{PhysRevLett.70.845,PhysRevLett.80.5631}. These deformations were also observed by local resistivity measurements \cite{PhysRevB.57.12781} and are consistent with those required for phase slip and CDW-to-normal carrier conversion at the contacts~\cite{PhysRevLett.70.845}. 
    
\begin{figure}[!ht]
    \centering
    \includegraphics[width=4.37in]{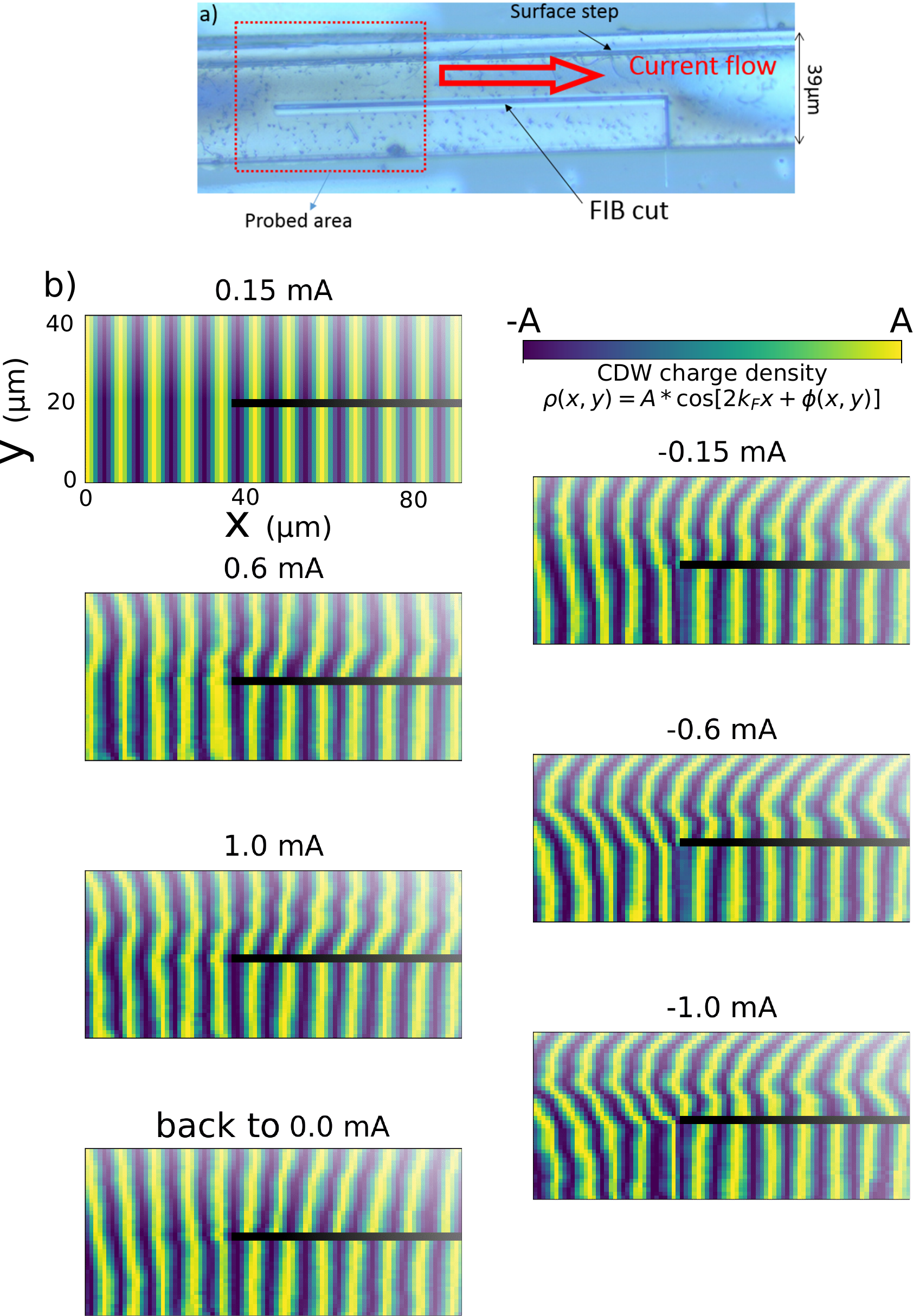}
    \caption{a) Image of the NbSe$_3$ wire. The focused ion beam (FIB) cut forces the current to flow through the upper part of the sample only, and not in the lower part below the cut line. This geometry allows to simultaneously observe sliding and non-sliding areas from a single sample. The CDW wave fronts are displayed in yellow. The area mapped by the x-ray beam is indicated in red. b) Evolution of the CDW in this region under increasing positive currents and decreasing currents down to negative values (I=\{0.15, 0.6, 1, 0, -0.15, -0.6, -1\} mA). The map at 0.15mA has been considered as the reference map. For clarity, the period of the CDW (yellow wave fronts) has been considerably increased (in reality, the CDW period is $\lambda$=14$\mathring{A}$) \cite{bellec2019evidence}).}
    \label{fig:ewen_shear}
    \end{figure}

     Although predicted in the 80's \cite{Feinberg}, at least in the close vicinity of the surface, the CDW shear deformation, which takes place along the direction perpendicular to current injection, had never been observed. This is explained by the fact that NbSe$_3$ samples have the shape of very thin wires, tens of micrometers wide requiring the use of a much smaller beam to observe transverse deformations without space averaging. 
      In this regard, an x-ray micro-diffraction experiment has been performed in NbSe$_3$ versus applied $dc$ currents. Four gold contacts were evaporated on a 39$\mu m\times$3$\mu m\times$2.25$mm$ single crystal glued on a sapphire substrate to perform four points resistivity measurements in-situ. Fast scanning diffraction technique allows us to map the CDW sliding across the NbSe$_3$ cross section. The precise 2k$_F$ wave vector has been measured as a function of the x-ray beam position on the sample surface. As shown in Fig.~\ref{fig:ewen_shear}, 100$\mu m\times40\mu m$ maps were probed with 1$\mu m$ resolution as a function of current. From these diffraction patterns, the CDW phase has been obtained by using a phase gradient method~\cite{bellec2019evidence}. The maps in Fig.~\ref{fig:ewen_shear} have been obtained using the gradient method by considering the map measured at I=0.15mA as the reference map. Indeed, due to hysteresis effects, it is always difficult to start the current injection from the true virgin state. However, we have considered that the reference map chosen was similar to the CDW initial state without current (see \cite{bellec2019evidence} for more details).

     In the upper part of the sample, in which current flows, a continuous deformation is observed from one lateral surface to the other, while the part in which no current flows remains unchanged. Like a guitar string plucked at both ends and subjected to a transverse force, the CDW bends in one direction or the other depending on the current direction.
     Despite the imperfections of the crystal, the CDW displays a continuous shear through the whole sample cross section, i.e. across 20$\mu m$, which corresponds to more than 10000 times its wavelength ($\lambda_{CDW}$=14\AA).
     This continuous deformation spreading over such a large distance, and leaving both boundaries unchanged, emphasizes that a CDW is able to maintain its cohesion over macroscopic distances despite the local disorder. A CDW, a least in NbSe$_3$, is mainly pinned by the lateral surfaces and little by the bulk, in contradiction with bulk pinning theories~\cite{Fukuyama1978}.
     
    Another indication that the surface has a dominant effect on CDW sliding is that the threshold current depends on the sample size and increases as it decreases. Resistivity measurements show that the threshold value diverges with decreasing sample length in NbSe$_3$ \cite{PhysRevB.32.2621,YETMAN1987201} and in TaS$_3$ \cite{MIHALY1983203}. Resistivity measurements have also shown that the threshold field is sensitive to the lateral dimensions of the sample, and increase with decreasing sample cross-section in NbSe$_3$ \cite{PhysRevB.46.4456,YETMAN1987201} but also in TaS$_3$ \cite{BORODIN198673}.
     
      To describe this sample length-dependence of the threshold, a phenomenological relationship between $E_{th}$ and sample length $L_x$ can be established, involving CDW bulk impurity pinning \cite{Feinberg}. Batistic {\it et al}. numerically found $E_{th}\approx 2.55 L_x^{-\alpha}$ where $\alpha = 1.23$ considering longitudinal pinning \cite{PhysRevB.32.2621}, but this, however, does not explain the constant $E_{th}$ observe for large $L_x$. On the other hand, a more precise description of the compression-dilatation profile developing along the CDW direction in NbSe$_3$ has been obtained by considering nucleation processes of dislocation loops considering creep effect and an incomplete conversion process \cite{brazo2000}. 

       \begin{figure}[!ht]
\includegraphics[width=6.37in]{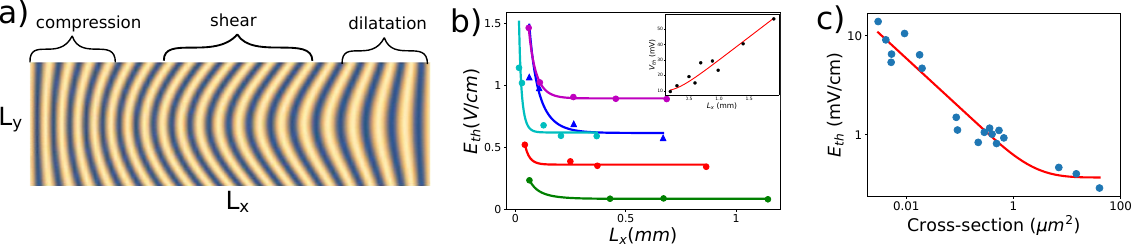}
\caption{a) CDW including the $\phi(x,y)$ obtained from Eq. ~\ref{phase_calcul}  displaying  a shear effect with a curvature of the wave fronts in the middle part of the sample and a compression-dilation of the CDW period at the two electrical contacts. The CDW wavelength $\lambda$ has been significantly increased for clarity (in reality $\lambda =14$\AA\ in NbSe$_3$, that is $\lambda \approx 10^{-6}L_x$) b) Threshold field $E_{th}$ versus $L_x$ and
its corresponding fit using equation (13) of Ref \cite{bellec2020}. The experimental dots were obtained in NbSe$_3$ (reproduced from \cite{PhysRevB.32.2621})  and the blue triangles in TaS$_3$  (from \cite{MIHALY1983203}). c) Threshold field $E_{th}$ (blue dots) versus sample cross section in small o-TaS$_3$ samples (reproduced from \cite{BORODIN198673}). The fit (red line) correctly reproduces the increase of $E_{th}$ for decreasing cross sections and the asymptotic constant value for large cross sections \cite{bellec2020}.}
\label{fig:bellec_threshold}
\end{figure}

     This experiment obviously highlights the predominant role of pinning by lateral surfaces, until now neglected in previous theories (see Fig.~\ref{fig:ewen_shear}).
     In order to get the spatial dependence of the phase and the size-dependent threshold field $E_{th}(L_x,L_y,L_z)$, let us consider the known 3D CDW free energy \cite{PhysRevB.48.4860,hayashi2000ginzburglandau}:
\begin{equation}
\mathcal{F}[\phi] \propto \int d^3\vec{r} \, \lbrace c_x^2 \phi_x^2+c^2_y \phi_y^2+c^2_z \phi_z^2  
+\omega_0^2\left[1-\cos(\phi)\right] +\eta Ex\phi_x \rbrace
\label{FreeEnergyPhase3D} 
\end{equation}
where $c_x,c_y,c_z$ are the CDW longitudinal and transverse elastic coefficients, $\phi_j\equiv \frac{\partial \phi}{\partial j}$ are the phase derivatives, $V_{imp}(\phi) \equiv \omega_0^2[1-\cos(\phi)]$ is a standard emulation of the bulk impurity pinning potential \cite{gruner2018density,Gork'ov1989} neglecting its randomness~\cite{Fukuyama1978,gruner2018density} and the last term corresponds to the CDW coupling to an applied electric field $E$ with the longitudinal gradient $\phi_x$, where $\eta$ is a temperature-dependent coupling coefficient \cite{hayashi2000ginzburglandau} and $Ex$ is the applied electric potential. Guided by the experiment, we fix the phase at the electrical
contact (x = 0 and x = Lx) and at the transverse surfaces. The corresponding boundary
conditions are :
\begin{equation}
\label{boundariesConditions}
\phi\left(\pm \frac{L_x}{2},y,z\right) = \phi\left(x,\pm \frac{L_y}{2},z\right) =  \phi\left(x,y,\pm \frac{L_z}{2}\right) = 0,
\end{equation}
The variational equation for the functional \ref{FreeEnergyPhase3D} (see Eq.~\ref{differentielle} below) was solved using the Green function and image charges method (see details in \cite{bellec2020,bellecphd2019}). In the first order in $\beta$, the solution yields:
\begin{equation}
\phi(\vec{r})\approx-\frac{32}{\pi^5}E\eta \beta \cos(\pi \frac{x}{L_x})\cos(\pi \frac{y}{L_y})\cos(\pi \frac{z}{L_z})
\label{phase_calcul}
\end{equation}
where the coefficient $\beta$ dependents on the sample size $L_x,L_y,L_z$ as:
$$\beta= \frac{1}{\frac{c_x^2}{L_x^2}+\frac{c_y^2}{L_y^2}+\frac{c_z^2}{L_z^2}+\frac{\omega_0^2}{2\pi^2}}$$
As shown in Fig.~\ref{fig:bellec_threshold}, Eq.~\ref{phase_calcul} for the phase satisfies the Dirichlet conditions (Eq.~\ref{boundariesConditions}). The solution corresponds to a CDW shear in the central part of the sample and a compression or a dilatation of the CDW wave fronts at the two edges, in agreement with experiments.

The threshold dependence on L$_x$ and L$_y$ can be obtained by considering a threshold strain $\phi'$ leading 
to $E_{th}\propto 1/L_{x}$ and to a constant $E_{th}$ at large $L_{x}$. Experimental data in Fig.~\ref{fig:bellec_threshold}.b) are shown with their corresponding fit using equation 13
from reference \cite{bellec2020}. The same equation was used to fit the evolution of $E_{th}$ as a function of
the sample cross-section S = L$_y$L$_z$ shown in Fig.~\ref{fig:bellec_threshold}.c). Furthermore, bulk impurity pinning
was removed for those fits ($\omega_0$=0) showing that surface pinning alone is sufficient to explain
the constant $E_{th}$ at large L$_x$ (see \cite{bellec2020} for more details).

As a conclusion, without considering the empirical bulk pinning  ($\omega_0=0$), and by only fixing the phase at a constant value on all surfaces, the global deformation of the CDW under current can be reproduced, including the dilatation and the compression close to electrical contacts\cite{PhysRevLett.70.845,PhysRevLett.80.5631}, the wave front curvature in the middle part \cite{bellec2019evidence}, and the threshold field dependence on sample length and cross section. 

This observation raises questions about the very nature of this phase able to develop a continuous deformation across macroscopic distances in imperfect lattices containing many defects in volume \cite{Bardeen:1979hv}. 
We can also note that if the transverse deformation is clearly related to lateral surface pinning, the longitudinal one isn't since the CDW compression-dilatation at the electrodes is also due to the conversion of normal electrons from the metallic electrode into  condensed charges~\cite{brazo2000,PhysRevLett.70.845}.

    \subsection{Sliding CDW based on a traveling soliton lattice.} 
    
In the previous chapters, we have shown several aspects of a CDW that are all related to sliding: CDW can stabilize dislocations at rest (see Fig~\ref{fig:fig3})~\cite{Bolloch2005}, those phase shifts are mobile above the threshold, leading to the disappearance of speckles in blue bronze (see Fig.~\ref{fig:fig5}) and in NbSe$_3$ (see Fig.~\ref{fig:nbse3})~\cite{PhysRevLett.109.256402}; the CDW displays a strong distortion (longitudinal and transverse) in NbSe$_3$ and an additional periodicity appears on top of the CDW when sliding in K$_{0.3}$MoO$_3$ blue bronze \cite{PhysRevLett.100.096403}. The excess of current observed in CDW systems above the threshold could be related to all these observations.

\begin{figure}[!ht]
\includegraphics[width=3.37in]{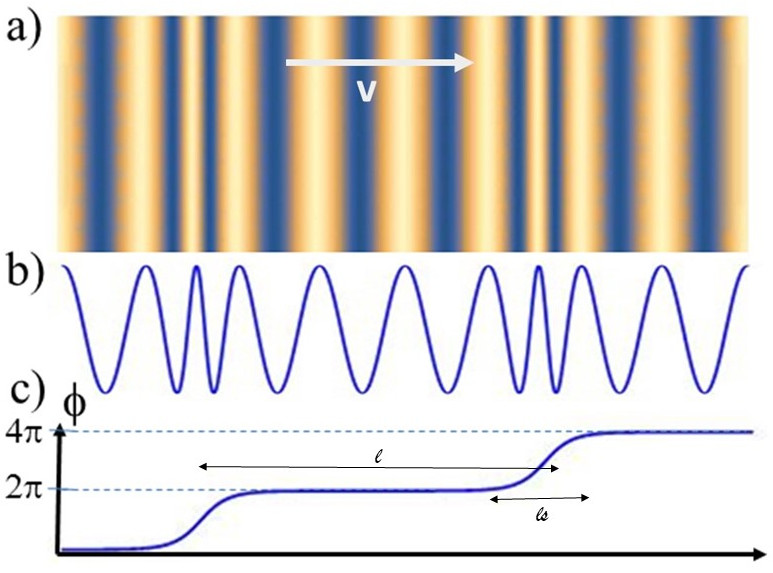}
\caption{Sketch of the soliton lattice in real space. (a) CDW in the presence of a soliton lattice with period $l$ corresponding to periodic 2$\pi$ phase shifts spread over a distance $l_s$ and propagating along the applied field (b) the corresponding electronic density and (c) the phase $\phi$ profile\cite{royo2016}.}
\label{fig:simulation0}
\end{figure}

Let us come back to the extra modulation observed in the sliding regime of the blue bronze (see Fig.~\ref{fig:secondary}). This modulation can be understood as the presence of a soliton lattice in translation (see Fig.~\ref{fig:simulation0}). For this, let us first consider a crude model considering the influence of defects through an interaction which couples the pinning potential and the phase $\phi$ \cite{Fukuyama1978}. The free energy leads to the following equation of motion in 1D~\cite{gruner2018density}:
\begin{equation}
\label{eq:theosol004}
\frac{\partial^2 \phi}{\partial t^2} - c_x^2 \frac{\partial^2 \phi}{\partial x^2} + \eta \frac{\partial \phi}{\partial t}+\omega_0^2 \sin \left( \phi \right) = F 
\label{differentielle}
\end{equation}
where $F=\frac{2 c_\phi^2 \, e \, E}{\hbar \, v_F}$ is proportional to the applied electric field,   $c_x = \sqrt{m / m^*} \, v_F$ is the phason's velocity and  $\omega_0$ the pinning frequency. We also add an effective damping term $ \eta \, \frac{\partial \phi}{\partial t}$ to mainly take  into account  the coupling between CDW and phonons. The $\sin \left( \phi \right)$ term is not linearized here allowing for abrupt phase variations. The usual non-perturbed sine-Gordon equation (for which F=$\eta$=0) is known to admit soliton solutions. However,  soliton excitations are quite robust and survive the inclusion of a reasonable external force and dissipation keeping their topological properties although the soliton shape is slightly modified \cite{PhysRevB.15.1578}. 
Let us now solve  Eq.~\ref{eq:theosol004} considering that the  phase $\phi(x,t)$ contains two terms: a slowly varying phase $\phi_0(x)$ and a dynamical part  $\phi_1(x,t)$ where $\phi_1$ varies much more rapidly than the static one. The static part $\phi_0(x)$ can then be calculated by  averaging Eq.~\ref{eq:theosol004} in time :
\begin{equation}
\label{eq:theosol008}
\left<  \frac{\partial^2 \phi_0(x)}{\partial x^2} \right>_t =(\frac{\eta \, \pi}{e} j- F)/c_\phi^2,
\end{equation}
leading to a quadratic variation of the phase $\phi_0(x)$   ($j = e/\pi \langle \partial \phi_1 / \partial t \rangle$).
The CDW, $\rho=\rho_0\cos(2k_F x+\phi(x,t))$, pinned at both ends at the metal/CDW junction, is compressed at one electrode and stretched at the other in agreement with experiments \cite{PhysRevLett.70.845,PhysRevLett.80.5631,brazo2000}.  
The excess of current  in the sliding regime $j =\frac{e}{\pi} \frac{\partial \phi}{\partial t}$ is constant  far from  electrodes as observed by numerous transport measurements~\cite{MAHER1993,gill1993, adelman1995, ITKIS:1997jf, Lemay:1998wx}.

\begin{figure}[!ht]
\includegraphics[width=3.37in]{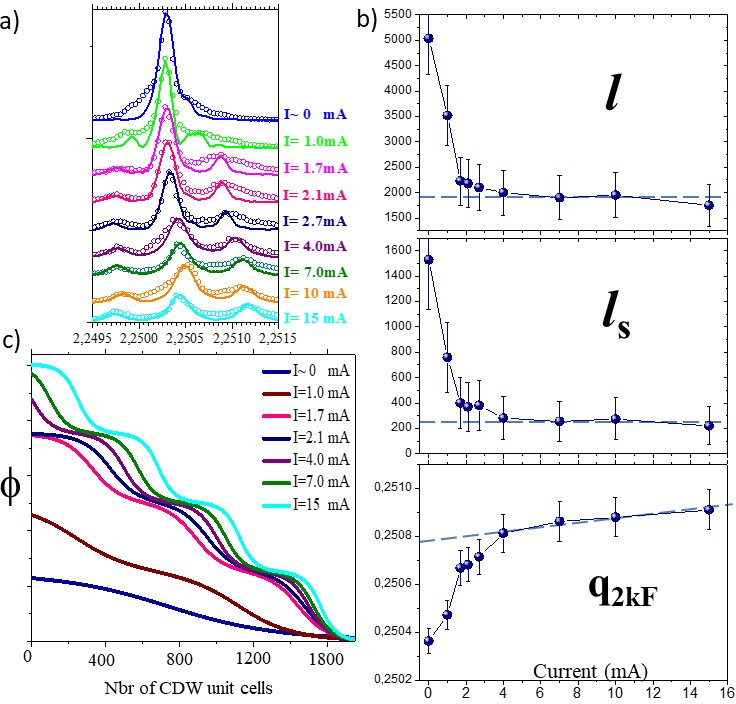}
\caption{a) Fit of the experimental data from the soliton lattice model (Eq.~\ref{eq:theosol019}). b) Fitted parameters $l_s$ and $l$ versus currents and b) profile of the corresponding phase $\phi$ showing an increase of the soliton density for increasing currents \cite{royo2016}.}
\label{fig:simulation}
\end{figure}

The dynamical part $\phi_1(x,t)$ obeys the sine-Gordon equation and is submitted to an effective force including friction.
Considering the periodic nucleation of CDW dislocations at the electrode~\cite{MAKI1995}, we obtain a train of solitons~\cite{PhysRevB.15.1578}:
\begin{equation}
\label{eq:theosol019}
\phi_1(x,t) = \delta+ \sum_{n=-\infty}^\infty 4 \arctan \left( \exp \left( \frac{x - v_S t - l n}{l_S \, \gamma (v)} \right) \right),
\end{equation}
where $l$ is the distance between successive solitons and $l_S= c_x/ \omega_0$ their extension (see Fig.~\ref{fig:simulation}). Overlapping effects between solitons are neglected ($l / l_S > 2$) \cite{royo2016}. Note also that the soliton lattice reaches a stationary sliding velocity $v_s$ proportional to the electric field E. Another expression for this soliton lattice, using the elliptic Jacobi function and giving essentially similar results, will be discussed later (see Eq.~\ref{Sol-SL}). 

This scenario of a soliton lattice traveling through macroscopic samples explains the main features of the collective transport of charges in CDW systems. In this approach, once created, these topological objects propagate without dispersion and with a remarkably long lifetime. Furthermore in this theory, the propagation velocity increases while the lattice period decreases with increasing applied fields and  these 2$\pi$ solitons carry a localized charge.  All these properties are in agreement with observations. It explains the well-defined frequency of the additional pulsed current despite the macroscopic dimensions of samples. Indeed, the charges remain spatially correlated despite the distance and the presence of disorder thanks to the robustness of these topological objects. The existence of the soliton lattice is also in agreement with the appearance of an additional periodicity observed by x-ray diffractioni experiments in the sliding regime of the blue bronze (see Fig.~\ref{fig:secondary}). It also explains why the fundamental frequency of the additional current increases with increasing currents  (since $v_s\propto$ E) and why the additional periodicity observed in the blue bronze decreases with increasing currents.

\section{Effects of solitons observed by diffraction}

In this chapter, some theoretical aspects of the physics of solitons will be described in relation to X-ray diffraction experiments. Only the unidirectional case, along the 2k$_F$ direction, is considered. 

CDWs are subject to stresses that can come from a variety of sources, such as surface effects inducing pinning and shear\cite{bellec2019evidence} and/or changes in transition temperatures \cite{brun2010surface}. CDWs can also be stressed by surface
doping, surface steps \cite{Isakovic06}, proximity to commensurability, various structural defects like twins, domain walls \cite{Rideau_2002}, a constraint geometry \cite{Latyshev} or an imbalance of normal and collective currents near junctions in the sliding regime \cite{brazo2000}. The stress can easily exceed an elastic limit leading to the appearance of topological defects. For commensurate or near commensurate CDWs, the associated stresses can leave particular fingerprints such as a soliton lattice or a system of random solitons \cite{Brazovskii2012,Kim_PhysRevLett.109.246802,Karpov2022,PhysRevLett.100.096403}. All these effects can, in principle, be observed by local probes such as STM or by space-resolved X-ray beam as previously described in this article (see also \cite{Rideau_2002,Isakovic:06,brazo2000,Lemay:1998wx}) although the interpretation is not always obvious.

The intensity of the elastic scattering of the CDW with a
distorted phase $\phi(\mathbf{r})$ is given by the expression:
\begin{equation}
I(\mathbf{q})=I_{0}\left|\int S(\mathbf{r}) \cdot e^{i\mathbf{q}\cdot \mathbf{r}}\cdot e^{i\left(\phi(\mathbf{r})-\phi(\mathbf{0})\right)} \cdot d\mathbf{r}\right|^{2}   \label{I(q)}
\end{equation}
where $I_{0}$ is the normalization constant, and $S(\mathbf{r})$ is
the correlation function describing an intrinsic disorder or inhomogeneity.
X-ray experiments usually recover the "square Lorentz" intensity profiles 
\cite{Ravy:2006} which correspond to a simple exponential decay in real space: $S(\mathbf{r})\sim\exp(-|\mathbf{r}|/\eta)$ where $\eta$ is the correlation length.

The $M$-fold commensurate CDW interacts with the underlying lattice via the
commensurability energy which, for a small CDW amplitude $A$, can be written
as (see e.g. \cite{gruner2018density}):
\begin{equation}
W_{com}=\alpha(1-\cos M\phi)~,~\alpha=A^{M}   \label{w-int}
\end{equation}
Cases $M=2,3,4,8$ are known in various CDW materials with $M=4$ for NbSe$_3$ and $M=8$ for Blue Bronzes; the case $M=1$ is to emulate the interaction with host impurities if we
ignore the randomness of their positions. With primary contributions to the
CDW energy $\varpropto A^{2}$, the commensurability effect is weak for $M>2$, affecting only the low energy deformations of the phase, leaving the amplitude $A$ almost unchanged.

For an exactly commensurate system, $W_{com}$ results in locking of the CDW
phase $\phi$ at values that are multiples of $2\pi/M$. The actual band
filling and/or local perturbations of the concentration of condensed
electrons enforce small deviations $q$ from commensurability. The overall phase displays
increments $\Delta\phi=qL$ over a length $L$ and the phase
jumps $N=\Delta\phi M/2\pi$ times along the sequence of commensurability
plateaux, forming a lattice of solitons (see experimental Fig.~\ref{fig:simulation}, and theoretical Fig.~\ref{Kir1} and Fig.~\ref{Kir5}). As previously described in the experimental section, the soliton lattice is characterized by two length
scales, the distance between solitons $l=\frac LN=2\pi/(Mq)$ and the soliton width $\xi$ (similar to $l_s$ in the previous arctangent model, see Eq.~\ref{eq:theosol019}). The period $l$ depends on band filling which makes it temperature-dependent due to thermal activation of carriers 
\cite{Pouget1991,Artemenko1996}. The single soliton width $\xi\varpropto(C_{x}/\alpha)^{-1/2}$ is jointly determined by the energy $\alpha$ and the CDW
rigidity parameter $C_{x}$. It also depends on the carriers concentration
due to the so-called effect of Coulomb hardening \cite{pouget91PRB,hennion1992,Artemenko1989, PhysRevB.69.115113}.
The basic model describing effects of the commensurability upon both the
ground state and the spatio-temporal evolution of the CDW phase relies upon
equation \cite{PhysRevB.15.1578}

\begin{equation}
m^{\ast}\frac{\partial^{2}\phi}{\partial t^{2}}+\gamma\frac{\partial\phi }{\partial t}-C_{x}\frac{\partial^{2}\phi}{\partial x^{2}}+\frac{e}{\pi }E^{\ast}+F_{com}=0~,~F_{com}(\phi)=\alpha M\sin\left( M\phi\right) 
\label{phi}
\end{equation}
whose parameters will be discussed later. The simplicity of the model and
availability of exact solutions (the pure sine-Gordon limit with $\gamma=0,E^{\ast}=0$) provoked a great amount of studies through decades (\cite{royo2016} and references therein). However, the precise measurement of the different parameters of this equation is far from being simple. The effective {\it mass density} $m^{\ast}$ is accessible by optical measurements though the parameter $\alpha$ which is related to the measurable pinning frequency $\omega_{pin}$ as $\alpha M/m^{\ast}=\omega_{pin}^{2}$. With regards to the damping coefficient $\gamma$, it is related to the measurable conductivity of the collective CDW sliding $\sigma_{c}$ as $\gamma=e^{2}/(s\sigma_{c})$ where $s$ is the area per chain. The definitions of $\sigma_{c}$, the stiffness $C_{x}$, and the effective driving field $E^{\ast}$ require a comprehensive approach to CDW physics, including such essential elements as Coulomb forces and normal carriers. The full microscopical approach with some applications was reviewed in \cite{Artemenko1989}, the transparent equations have been derived in \cite{annals:2019} and the reasonably simplified version can be written as:

\begin{equation}
\frac{1}{\pi}\frac{1}{s\sigma_{c}}{\partial}_{t}\phi-{\frac{1}{\pi}{{\rho}_{c}/{\rho}_{n}{\partial}_{x}^{2}\phi\ +}}\frac{~F_{com}}{e^{2}N_{F}}=-\frac{J(t)}{s{\sigma}_{n}}~,~\frac{1}{\sigma_{c}}=\frac{1}{{\sigma}_{CDW}}+\frac{1}{{\sigma}_{n}}   \label{J(t)}
\end{equation}
which is comparable, term by term, to Eq.~\ref{phi} allowing to
interpret its components. The microscopically-derived temperature-dependent
parameters are the normal and condensed densities $\rho_{n}$ and $\rho_{c}=1-\rho_{n}$, with $\rho_{c}(T_{c})=0$. They are normalized to $\rho_{c}(0)=1$,
and hence $\rho_{n}(0)=0$. The second term on the left-hand-side describes the enhancement of the
phase elasticity $C_{x}$ and the ratio $\rho_{c}/\rho_{n}$ controls the rigidity vanishing, while approaching the metallic state at $T\rightarrow T_{c}$. It also explains the Coulomb hardening between condensed charges
 which dramatically increases with freezing out of screening by
normal carriers when $\rho_{n}\rightarrow0$ at $T\rightarrow0$. The right-hand-side of
Eq.~\ref{J(t)} gives the driving force $E^{\ast}$ as the total monitored
current ${J(t)}$ divided by the normal conductivity ${\sigma}_{n}$ alone. 
The total current is additive: $J=J_{n}+J_{CDW}$ with the intrinsic CDW
current (per chain) being $J_{CDW}^{1}=-(e/\pi)\partial_{t}\phi$.

These complications, not quite intuitive, arise from the complex interplay of
Coulomb interactions with screening facility of normal carriers. The
effects are particularly strong for solitons which are actually the walls crossing the sample; their charge density generates a constant electric field which must be screened by the cloud of normal electrons whose width is the screening length $\varpropto1/\rho_{n}$. This cloud deforms or moves together with the phase evolution thus contributing to both elastic energy and friction.

The boundary conditions applied to the solution of Eq.~\ref{phi} also require special attention. It is already known that without pinning, the solution in the dissipative limit is:
$\phi=(px^{2}-rt)$ where $p$ and $r$ are related as follows: $C_{x}p+\gamma r=eE^{\ast
}/\pi$. In the opposite case, without boundary conditions, there is a freedom to redistribute the action of the
driving force among the elastic $\sim p$ and viscous $\sim$ r contributions.
We will demonstrate below the difference in behavior under different boundary conditions, both physically motivated.

Experiments performed on CDW systems show that the sliding motion is
essentially dissipative. According to optical data (see e.g. \cite{gruner2018density}), the CDW response is overdamped for frequencies below $10GHz$, while the characteristic frequencies of CDW sliding, as measured by Narrow Band Noise (NBN) measurements, do not exceed $100MHz$. We must therefore keep only the derivatives in the dynamical equation (Eq.~\ref{phi}). The resulting equations can only be studied numerically, and is presented in the following.

The important question that arises is why the damping coefficient $\gamma$ is so large, as shwon by the experiments dealing with 
the interaction of CDW with phonons. The answer also deserves to be integrated
in the context of the physics of solitons, a now local phenomenon, caused by microscopic defects. Indeed, their presence is known to destroy the long-range CDW order. However, more importantly for us here, they provide an evolving metastable order, whose relaxation dissipates the sliding energy.

For the so-called local or strong pinning, the answer
is explicitly known (see \cite{Nattermann2004}). For small velocities, it looks like

\begin{equation}
\gamma=4n_{i}F_{\pi}\tau_{\pi}~,~F_{\pi}=\frac{1}{2}\left. {\frac{d\Delta E}{d\theta}}\right\vert _{\pi}
\end{equation}
where $n_{i}$ is the linear concentration of impurities, $\tau_{\pi}$ is the
relaxation time of the local metastable state, $F_{\pi}$ is the phase
restoring force produced by the impurity, $\Delta E$ is the energy term
dependent on the phase mismatch $\theta$, and $\tau_{\pi}$ is the relaxation
rate of the metastable state. More details are given in the last section.

In the following, we will first consider randomly-distributed static solitons, then a regular lattice of solitons, either static or moving under the effect of an external electric field, and finally the solitons in space and time domains appearing in the course of phase slips.

\subsection{Rare random solitons}

Consider a CDW system containing few solitons, with a mean concentration $c=1/l$,
trapped by impurities or thermally-scattered at $T>T_{c}$, so that their
positions $x_{i}$ are random. Each soliton produces a phase shift $\chi
=2\pi/M$. We shall keep $\chi$ to be arbitrary allowing also to include the effect of 
charged impurities, providing Friedel phase shifts even in absence of
solitons (see \cite{Ravy:2006} and refs. therein). The correlation function
in (\ref{I(q)}) can be calculated via the Poisson distribution for a number 
 of defects $n$ found within the interval $(0,x)$. We have

\begin{equation}
D(x) = \left\langle e^{\imath\left[\phi(x)-\phi(0)\right]}\right\rangle\\
= \sum_{n}e^{i\chi\cdot n\cdot sgn(x)}\frac{\left(c\left|x\right|\right)^{n}}{n!}e^{-c|x|}\\
= e^{-|x|c(1-\cos\chi)}e^{ixc\sin\chi}.
\end{equation}\\
In $q$ space, we find a single symmetrical Lorentzian peak with the width $c(1-\cos\chi)$ and located at the displaced position $\delta q\mathbf{=}c\sin\chi$:
\begin{equation}
I(q)=R\frac{2c\left(1-\cos\chi\right)}{\left(q-c\sin\chi\right)
^2+4c^2\left(1-\cos\chi\right)^2}\ \ ; \ \chi=\frac{2\pi}{M}
\end{equation}
where $R$ is the material parameter. If $\chi\ll1$, the shift is just given by the mean stretching $\mathbf{\delta q\approx}c\chi$. Even small, the shift is nevertheless visible because the
broadening $c\chi^{2}$ is even smaller. On the contrary, when the shift is
maximal ($\delta q=c$ at $\chi=\pi/2$ for $M=4$), for the unitary
limit of the impurity potential, the peak position is no longer well-defined since its broadening is of the same order of magnitude as the shift.

\subsection{Static lattice of solitons}

Without driving force, $E^{\ast}=0$, the static soliton of Eq.~\ref{phi}
yields the regular soliton lattice whose form is known analytically:

\begin{equation}
\phi(x,k)=\frac{2}{M}am\left[ \frac{M}{2k}\frac{x}{\xi},k\right] ~;~ l=\frac{4kK(k)\xi}{M} ~;~ \xi=\left( \frac{C_{x}}{\alpha}\right) ^{1/2} 
\label{Sol-SL}
\end{equation}
where $l$ is the distance between solitons, $\xi$ is the characteristic
soliton size, $am[\tau,k]$ is the elliptic Jacobi function, and $K(k)$ the
complete elliptic integral of the first kind. This expression is more precise than the one used previously (see Eq.~\ref{eq:theosol019}), although it may be less intuitive. 

\begin{figure}[h]
\includegraphics[width=5.37in]{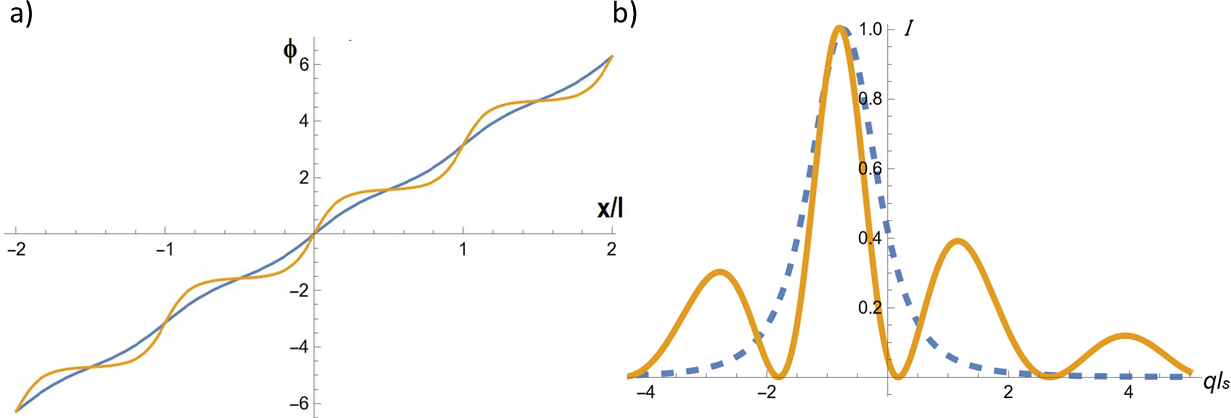} 
\caption{a) The phase $\phi$ grows almost linearly for $\xi=0.44l$ (blue line) or displays abrupt phase shifts $\xi =0.04l$ (yellow line) b) Corresponding normalized diffraction profile of the sattelite reflection associated to the CDW displaying the two types of solitonic lattice for $\xi=0.44l$ (dashed blue line) and for $\xi =0.04l$ (yellow line). The shift of the maximum of intensity is related to the average slope of the phase, which is equal in both cases here.}
\label{Kir1}
\end{figure}

The intensity $I(q)$ is determined by Eq.~\ref{I(q)} containing the phase $\phi$ given by Eq.~\ref{Sol-SL}. The results are shown in Fig.~\ref{Kir1}. When the soliton size is large, for $\xi\sim l$ ($k\ll1
$), the phase $\phi$ grows almost linearly. In that case, the intensity $I(q)$ is simply
shifted in q-space from the commensurability point $Q_{0}$ to $Q_{0}+2\pi/l$. In the opposite strongly non-linear case, where $\phi$ contains abrupt phase shifts ($\xi<<l$), the solitonic superstructure leads to the formation of
 non-symmetric peaks, spaced by a wave vector depending of $1/l$ (Fig.~\ref{Kir1}).

\subsection{Lattice of solitons submitted to an external electric field}

Consider now the static and the stationary states of the solitonic lattice
under the applied homogeneous field $E^{\ast}.$ Recall that the CDW sliding is overdamped, and so we consider the behavior of the phase $\phi$ described by  Eq.~\ref{phi} in the limit $m^{\ast}=0$. We numerically solve the differential equation by considering two different physical situations: 
\newline
(i) The CDW condensate is considered as isolated, with a conserved total charge and submitted to
boundary conditions fixing the phase increment:
\begin{equation}
\Delta\phi=\phi(L,t)-\phi(0,t)=2\pi N_{s}/M   \label{BCa}
\end{equation}
where $N_{s}$ is the number of solitons over the chain length $L$ present
before the field is applied.
\newline
(ii) The CDW condensate without boundary conditions, only the initial phase distribution $\phi(x,0)$ is defined.

Independently of the boundary conditions, the CDW state is not static anymore
for too large $E$, exceeding a critical electric field $E_{c}$. In the specific case of the rigid approximation, where $\partial_{x}\phi=0
$, the potential energy $-\alpha\cos(M\phi)-eEx/\pi$ behind the Eq.~\ref{phi} looses its minima above the critical electric field $E_{c0}=M\alpha$. In the more general case, where $\partial_{x}\phi\neq0$, the threshold field $E_{c}$ is different and our numerical
solution shows that the actual threshold field is lower than the rigid case, with $E_{c}\approx0.7E_{c0}$, being reduced by the allowed elasticity $\partial_{x}\phi\neq0$.

\begin{figure}[ptb]
\includegraphics[width=\linewidth]{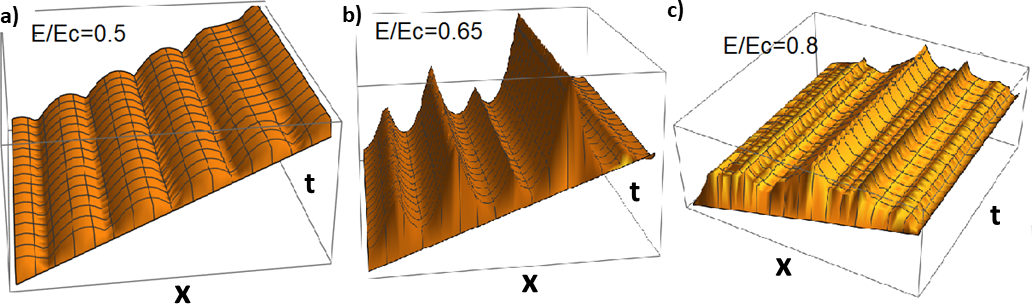} 
\caption{Space-time distribution of the phase $\protect\phi$ for various  electric fields: a) $E=0.5E_{c}$, b) $E=0.65E_{c}$ and c) $E=0.8E_{c}$.}
\label{Kir2}
\end{figure}

\begin{figure}[ptb]
\includegraphics[width=\linewidth]{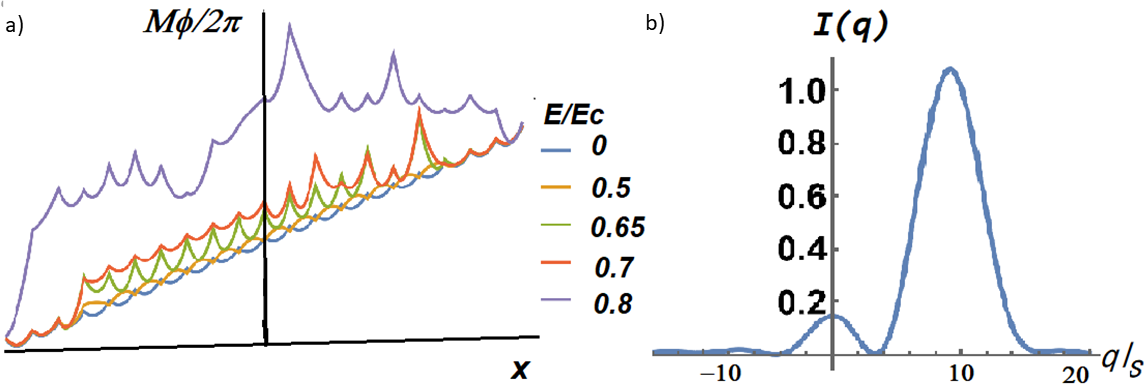} 
\caption{a) Space distribution of the phase $\protect\phi(x)$ at a fixed
time for various electric fields. b) Scattering intensity profile for an
intermediate field $E/E_{c}$=0.5.}
\label{Kir3}
\end{figure}

For the case (i) of the closed system, the space-time distribution of the
phase $\phi(x,t)$ is presented in Fig.~\ref{Kir2} for low and high electric
fields. 
We see that the static solution corresponding to the lattice of somehow
deformed solitons is reached at sufficiently long times. At the critical
field, the soliton lattice starts to move and no static limit can be reached.
Fig.~\ref{Kir3}a shows the space dependence at a given point $%
t_{0}$ of $\phi(x,t_{0})$ for various electric fields. In the sliding regime,
the amplitude of undulations diminishes. At larger $E$, the profile of $%
\phi(x,t_{0})$ approaches a plateau while the solitons are expelled in favor of a steep rise of $\phi$ near a boundary. The scattering intensity $I(q)$ is presented in Fig.~\ref{Kir3}b for a subcritical electric field.

\begin{figure}[ptb]
\includegraphics[width=4.37in]{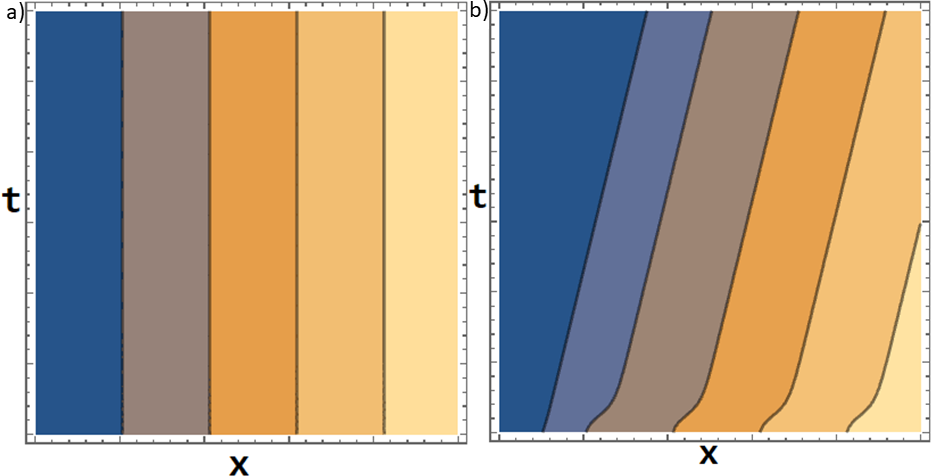} 
\caption{Contour plots of phase $\protect\phi$ for a) $E=0$ and b) $E=0.15E_{c}$.}
\label{Kir4}
\end{figure}

\begin{figure}[ptb]
\includegraphics[width=\linewidth]{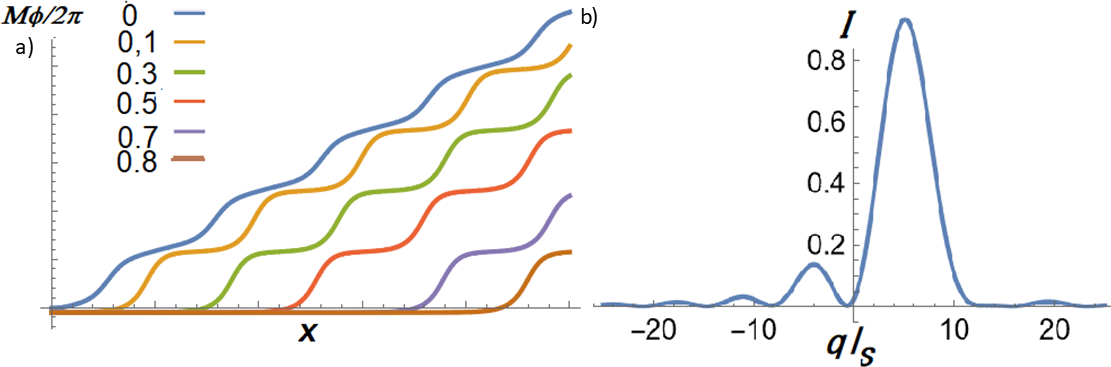}
\caption{a) Space distribution of the phase $\phi(x)$ at a
fixed time for various electric fields. b) Scattering intensity profile
for intermediate field.} 
\label{Kir5}
\end{figure}

For the case (ii) of the open system, we imply the initial conditions
corresponding to the initial solitonic distribution: 
\begin{equation}
\phi(x,0)=am(x,k) 
\end{equation}
The results of the numerical solution are presented in Fig.~\ref{Kir4}
and Fig.~\ref{Kir5}. At short times, the soliton lattice profile $\phi(x,t)$ changes and becomes non-symmetrical. The soliton width increases with increasing applied fields, and the solitons start moving. The density plot of $\phi(x,t)$ is presented in Fig.~\ref{Kir4}.

At zero electric field (Fig.~\ref{Kir4}a), the initial periodic profile $\phi(x,0)$ 
is kept statically, but it starts to move under applied external electric
fields with no threshold. The contour lines $\phi(x,t)$ bend and the distance
between solitons changes not only in space but also in time (Fig.~\ref{Kir4}b). The behavior of the phase in the middle part of the sample, $\phi(L/2,t)$, for various
electric fields is presented in Fig.~\ref{Kir5}a; the intensity $I(q)$ at an intermediate electric field is presented in Fig.~\ref{Kir5}b.

In this complex CDW sliding process, many questions can be answered by considering the phase-only framework, like how
solitons enter or leave the sample, how they are (re)created near junctions, how the
total soliton number $N_{s}$ changes with temperature for example, or how the equilibrium wave vector evolves (see 
\cite{Zybtsev2016} and the first part of this review concerning diffraction experiments). However, a change of $N_{s}$, or more generally of the total phase increment $\Delta\phi$, is required for so-called phase-slip processes, which correspond to a kind of space-time vortices. As a consequence, it is required for the amplitude $A$ of the CDW order parameter $\Psi=A\exp(i\phi)$ to
vanish in the vortex core \cite{bellec2020}. Thus, the equations must be generalized and include
the function $A(x,t)$. The commensurability energy has to be generalized as $W_{com}\varpropto-\alpha\left( \Psi^{M}+\Psi^{\ast M}\right) $ and the following additional equation for $A$ has to be considered:
\begin{equation}
{{\kappa}_{x}\partial}_{x}^{2}A+{\kappa}_{x}A{\left( {\partial}%
_{x}\phi\right) }^{2}-A+A^{3}-\alpha M\left( \Psi^{M-1}+\Psi^{\ast
M-1}\right) =\tau^{-1}{\partial}_{t}A   \label{A}
\end{equation}
where $\kappa_{x}$ and $\tau$ (the amplitude relaxation rate) are some
constants. In Eq.~\ref{phi}, we must take into account that $\gamma\varpropto C_x\varpropto A^2$ and $\alpha\varpropto A^M$. The
vorticity can be obtained only in invariant variables, so the phase
derivative in Eqs.(\ref{phi},\ref{A}) must be generalized as $\partial_t\phi\rightarrow \omega=Im(\Psi^{\ast}\partial_t\Psi)/A^2$, $\partial_x\Psi\rightarrow q=Im(\Psi^{\ast}\partial_x\Psi)/A^2$ with the phase being restored as $\phi=\int^t\omega(t^\prime)dt^\prime$. 
The numerical solution was performed in terms of components $u,v$
of $\Psi=u+iv$. The boundary conditions are given in terms of $q$ ; in view
of the local electroneutrality condition $(q/\pi)+n=0$, $q$ specifies the
concentration of normal electrons $n$, and thus their chemical potential which
is the standard assumption.

\begin{figure}[ptb]
\includegraphics[width=1\linewidth]{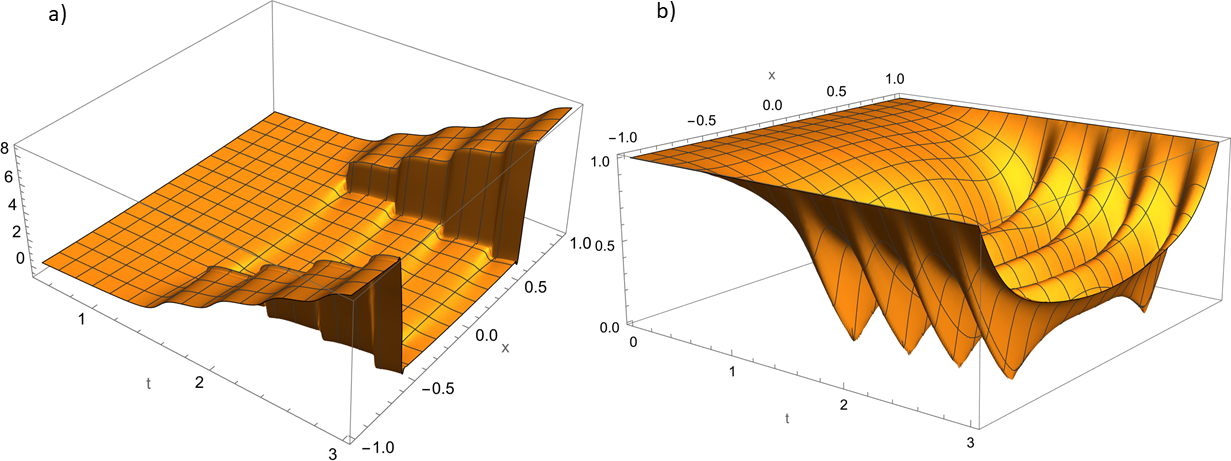} 
\caption{Phase slip processes. a) Space-time distribution of the phase $
\phi(x)$ and b) of the amplitude A(x,t).}
\label{Kir6}
\end{figure}

Fig.~\ref{Kir6} shows an example of numerical solutions for $M=1$. We see
a stratification among the pinned bulk where the phase is nearly constant
and the sliding stripes near junctions where the phase evolves by $2\pi$
pulses (solitons in the time domain). The regions are separated by a
periodic array of vortices in time, and form as a
wall in the $x$ direction, which can be viewed as space-domain solitons. The plot of the amplitude
shows the sequence of nodes: as expected, $A(x,t)$ goes to zero at
the space-time vortex centers.

In conclusion, the commensurability solitons can be observable in CDWs with
a sufficient concentration of normal electrons. Particular manifestations
near junctions are challenging for space-resolved studies, particularly
coherent micro-diffraction. These intriguing spatial and temporal effects require the use of both space and time-resolved techniques to be observed.

\subsection{Generation of pairs of solitons by an impurity and
resulting CDW viscosity.}

Consider a system of interacting CDW chains with a point
impurity located at position $\vec{r}_{i}$, $x=x_{i}$ on the chain $n=0$. We can write the
Hamiltonian as

\begin{equation}
H=\int dx\left\{ \sum_{n}\right[ \frac{1}{2}C_{\parallel}(\partial
_{x}\phi_{n})^{2}- \sum_{m}C_{\bot}\cos(\phi_{n}-\phi _{m})\left] -V\cos(\phi_{0}+\vec{Q}\vec{r}_{i})\delta(x-x_{i})\right\} 
\label{quasi}
\end{equation}
where $C_{\bot}$ is the interchain coupling and $V$ is the impurity
strength. The $2\pi$ periodicity of the pinning energy allows to skip the $2\pi$ quanta in $\phi_{0}$ to optimize the total energy. Moreover, the $2\pi$ periodicity of the regular energy in Eq.~\ref{quasi} allows for
interchain $\pm2\pi$ solitons. For the soliton centered at position $X$,
the phase profile $\phi_{s}(x-X)$ describes stretching/dilatation by one
period along the defected chain relative to the surrounding ones. The soliton is
distributed over the length $\xi\sim\sqrt{C_{\parallel}/C_{\bot}}$\ and
costs the energy $E_{s}\sim\sqrt{C_{\parallel}C_{\bot}}$, the two terms defining
the equilibrium concentration of solitons $n_{s}\sim\exp(-E_{s}/T)$.

The energy should be minimized over $\phi(x)$ with the asymptotic
condition $\phi\rightarrow\bar{\phi}$ at $|\mathbf{{x}-{x}_{i}|\rightarrow\infty}$ where the mean phase in the bulk $\bar{\phi}$ can be time-dependent. It is convenient to keep the local value $\phi_0(\mathbf{{x}_{i})}$ fixed and optimize it only at the end of the calculation. Then the pinning center can be
described by a single degree of freedom $\psi_{i}$ and monitored by another
single one $\theta_{i}$. Let us define the local mismatches of phases relative to the bulk value $\bar{\phi}$:
\begin{equation}
\psi_{i}=\phi(\mathbf{{x}_{i})-\bar{\phi}\;,\;\theta_{i}=-{Qx}_{i}-\bar{\phi}~,\partial_{t}\bar{\phi}=\omega}
\end{equation}
 Henceforth, the index $i$ will be omitted.

\begin{figure}[ptb]
\includegraphics[width=1.\linewidth]{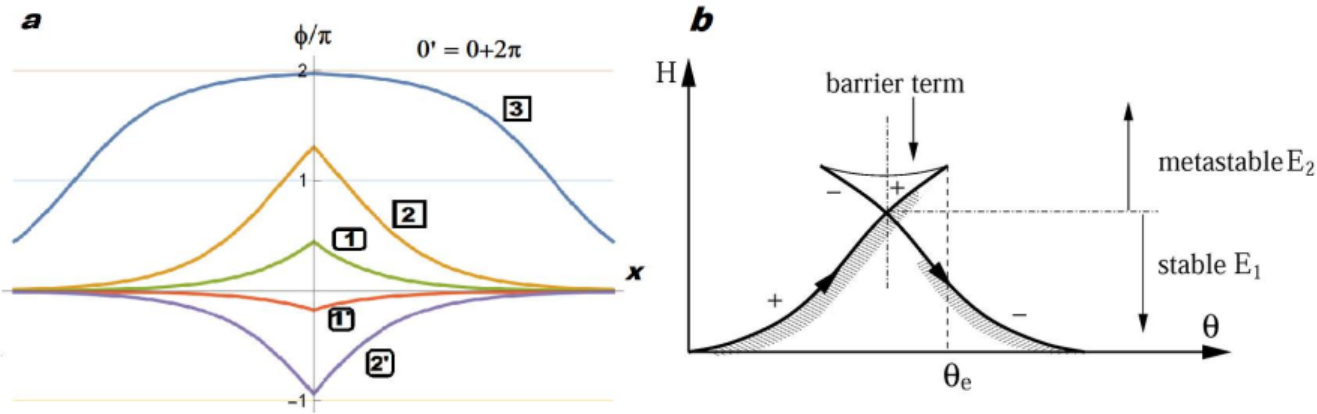} \label{branches} 
\caption{a) The $\phi$ profile starts from the equilibrium position $0$ with
$\phi(x)\equiv 0$, and evolves through the shapes $1,2,3, 0^\prime$. These
configurations correspond to the retarded branch $E_{+}$ which becomes
metastable after $\protect\phi(0)$ crosses $\protect\pi$. The phase $\phi$ will then follow the
advanced profiles $1^{\prime},2^{\prime},3^{\prime},0$, corresponding to the branch $E_{\_}$. If the retarded branch $E_{+}$ is less deformed, costing a smaller energy $W$, the relaxation $E_{+}\rightarrow E_{\_}$ is avoided, and the new development starts with the
profile $0^\prime=0+2\pi$ corresponding to the infinitely divergent
pair of solitons. b) Energy branches for a bistable
impurity. The upper line shows the barrier branch $E_{3}$. Solid
lines show the locally stable branches $E_{\pm}$, also classified as $E_{2}>E_{1}$. The difference $\Delta E=E_{2}-E_{1}$ gives the dissipated
energy. The difference $U=E_{3}-E_{2}$ gives the activation energy for a
decay of the metastable state $E_{2}$.}
\end{figure}

Quantitative results can be obtained within a short-range model (Eq.~\ref{quasi}). If we consider that only the central chain $n=0$ (passing through the
impurity) is perturbed while its neighbors stay at $\phi_{n\neq0}\equiv 
\bar{\phi}$ homogeneously, then the energy functional can be simplified as: 
\begin{equation}
\int dx\left[ \frac{1}{2}C_{\parallel}(\phi^{\prime})^{2}-C_{\bot}\cos
(\phi)-V\cos(\phi-\theta)\delta(x)\right]   \label{srm}
\end{equation}
Its extremum is the function $\phi(x)=\phi_{s}(x-X)-\phi_{s}(x+X)$
where $\phi_{s}(x)$ is the standard sine-Gordon soliton shape and $X$ is
fixed by the conditions $\phi(0)=2\phi_{s}(X)=\psi$. The successive 
 $\phi(x)$ profiles versus $\theta$\ is shown in Fig.~\ref{branches}a. 

The energy can be written as: $W(\psi)=E_{s}(1-\cos (\psi/2))$.
Over one period, $W(\psi)$ changes monotonously within $0=W(0)\leq
W(\psi)\leq W(2\pi)=2E_{s}$. The remnant variational energy contains the
pinning potential $V(\psi-\theta)$, which we take as $V(\phi)=V(1-\cos\phi)$, and the energy of deformations $W(\psi)$ with $W(0)=0$, $W(2\pi)=2E_{s}$: 
\begin{equation}
H(\psi,\theta)=V(\psi-\theta)+W(\psi)   \label{H=W+V}
\end{equation}
The study of the extrema of this energy yields one or three solutions $\psi_{i}(\theta),~i=1,2,3$ whose energies $E_{i}(\theta)$ are illustrated in Fig.~\ref{branches}b. As an example, the profile $i=3$ corresponds to a dilatation of the CDW wavelength at one side and a compression on the other side, in agreement with the observations~\cite{PhysRevLett.70.845,PhysRevLett.80.5631}. The whole interval of $\theta$, or some parts of it, can be
either \textit{mono-stable} or \textit{bi-stable}. The last case corresponds
to the coexistence of two locally stable branches: the \textit{absolutely
stable} one with the lower energy $E_{1}$ and the \textit{metastable} one with
a higher energy $E_{2}$. The same pair of branches can be regrouped also as
the \textit{ascending} branch $E_{+}$ for which $F_{+}(\theta)>0$ and the 
\textit{descending} one $E_{-}$ with $F_{-}(\theta)<0$, where $E_a$ (with $a=\pm$) are the partial forces generated by the pinned state $a$. They correspond to
the \textit{retarded} and the \textit{advanced} states at the impurity,
respectively, and the two branches cross each other at $\theta=\pi$, with $\psi_{\pm}(\pi)<\pi $ (see Fig.~\ref{branches}b). The barrier height, with
respect to the metastable branch $E_{2}$, gives the activation energy for its
decay: $U_{b}(\theta )=E_{3}(\theta)-E_{2}(\theta)$.

Let us consider the stationary process when the CDW moves with a constant
phase velocity $\omega=-\dot{\bar{\phi}}=\dot{\theta}=const$. The pinning
force can be written as a weighted distribution of instantaneous forces: 
\begin{equation}
F_{pin}=n_{i}\int_{\pi}^{\theta_{max}}d\theta{F(\theta)}\exp\left( {-\int_{\pi}^{\theta}{\frac{d\theta_{1}}{\omega\tau(\theta_{1})}}}\right)
\,\,,\;F={\frac{d}{d\theta}}\frac{\Delta E}{2}=\frac{F_{+}-F_{-}}{2} 
\label{f-the-F}
\end{equation}
where the expression in the exponent generalizes, for a variable relaxation time $\tau
(\theta)$, the natural guess for the decay probability as $\exp(-T_{slide}/\tau)$ where $T_{slide}=2\pi/\omega$ is the period of the
CDW sliding over the impurity site. 

We shall limit the discussion to small
velocities $\omega\ll\tau_{\pi}^{-1}$ where $\tau_{\pi}=\tau(\pi)\sim\exp(U_{\pi}/T)$, $U_{\pi}=U(\pi)$, is the maximal relaxation time in the
region of the branch crossing the point $\theta=\pi$. The main contribution
comes from the close vicinity of $\pi$: $\theta\approx\pi+\delta\theta$
where $\delta\theta \sim\omega\tau_{\pi}$. We can distinguish between two
sub-regimes.

1. \emph{Very small velocities}: with $\omega\ll
\omega_{\pi}=T/(\tau_{\pi}F_{\pi})\ll\tau_{\pi}^{-1}$ and $F_{\pi}=F(\pi)$. The
decay happens as soon as the branch becomes metastable in a vicinity of $\pi$, even before the $\theta$ dependence is seen. The life time
interval is $\delta\theta\sim v\tau_{\pi}$, hence Eq.~\ref{f-the-F} yields
the expression 
\begin{equation}
F_{pin}=n_{i}\omega F_{\pi}\tau_{\pi}   \label{tau-pi}
\end{equation}
which gives the phenomenological viscosity in the regime of the \textit{\ 
linear collective conductivity}. It shows an activated behavior via $\tau_{\pi}^{-1}$ which can emulate the normal conductivity via thermally
activated quasi-particles.

2. \emph{Moderately small velocities}:
with  $\omega_{\pi}\ll\omega\ll\tau_{\pi}^{-1}$ and 
\begin{equation}
F_{pin}\sim n_{i}T\ln[\omega\tau_{\pi}]\,\quad \textrm{i.e.} \quad\omega\sim\tau_{\pi}^{-1}\exp(f/n_{i}T)  
\end{equation}
Convenient interpolation formulas for the two cases 1,2 can be obtained: 
\begin{equation}
F_{pin}\approx Tn_{i}\ln\left( 1+\omega\frac{\tau_{\pi}F_{\pi}}{2\pi T}\right) ; \quad\omega=\frac{2\pi T}{\tau_{\pi}F_{\pi}}\left( \exp \frac{F_{pin}}{Tn_{i}}-1\right) 
\end{equation}

The physics of $F_{pin}\sim\omega$ regime is given by the high probability to stay
with the metastable branch in the course of  small displacements $\delta\theta\sim$ $\tau_{\pi}v$. The  $F_{pin}\sim\ln\omega$ regime appears because that at higher $\omega$ a wider region of $\delta\theta$ is explored and the metastable
branch starts to feel the decrease of the barrier (long in advance, there is
either the termination point $\theta_{e}$ or the minimal barrier point $\theta_{m}$), even if still unreachable at these moderate $\omega$. The complete
range of velocities was described in \cite{Nattermann2004} and compared with
experiments in \cite{Ogawa2005}.

\subsection{Conclusion}
This article summarizes different studies that have made it possible to observe defects in electronic crystals, such as charge and spin density waves, thanks to cutting-edge X-ray diffraction techniques and more particularly coherent X-ray diffraction. New information about the CDW static and dynamical regimes could be obtained thanks to the sensitivity of coherent x-rays to phase variations, especially in the sliding CDW state. 
Two main features emerge from this series of experiments: the CDW deformation, mainly  due to pinning by sample surfaces and the propagation of a periodic lattice of phase shifts on top of the CDW. These two characteristics of sliding CDW are linked. The collective transport of charges in CDW systems is based on the dynamics of phase shifts induced by CDW deformations. After a threshold strain, the CDW elastic energy is released by creating a 2$\pi$ soliton. Once created, the soliton propagates freely across the sample. Since the applied current is continuous, the  creation of soliton is periodic, generating a soliton lattice in motion. 

There is, however, still a lot to understand. The main difficulty lies in the fact that each system, despite identical aspects, has different properties. The diversity of measured responses  probably shows that the type of transport is not identical depending on the system. Ultrashort and coherent x-ray pulses as the ones delivered by X-ray Free Electron Lasers could open new perspectives thanks to unprecedented spatial and temporal resolution to get a better understanding of this collective motion of charges.

\subsection{Acknowledgements}
The work summarized in this review spans over 20 years and involves a large number of researchers. Without being able to name them all, we would first like to thank the PhD students who played an essential role in these works, E. Pinsolle and A. Rojo-Bravo. We would also like to give special thanks to A. Sinchenko whose mastery of resistivity and sample preparation techniques has always been the basis of all these experiments. We also thank all the collaborators who accompanied us (S. Ravy, C. Laulhé), and more particularly P. Monceau for his precious advice and for providing us with samples and, from a theoretical point of view, S. Brazovskii, especially for the last section of this review.

\bibliography{review-phase-final-archiv}

\end{document}